\documentclass[acmart]{acmart}

\usepackage{colortbl}  
\usepackage{xcolor}
\usepackage{url}
\usepackage{multirow}
\usepackage{subcaption}
\usepackage{arydshln}
\usepackage{tabularx}
\usepackage{enumitem}
\usepackage{array}   

\title{A Bi-Step Grounding Paradigm for Large Language Models in Recommendation Systems}

\newcommand{\eg}{\emph{e.g., }}

\begin{document}

\author{Keqin Bao*}
\email{baokq@mail.ustc.edu.cn}
\affiliation{%
  \institution{University of Science and Technology of China}
  \country{China}
}

\author{Jizhi Zhang*}
\email{cdzhangjizhi@mail.ustc.edu.cn}
\affiliation{
  \institution{University of Science and Technology of China}
  \country{China}
}

\author{Wenjie Wang}{
\email{wenjiewang96@gmail.com}
\affiliation{
  \institution{National University of Singapore}
    \country{Singapore}
}
}

\author{Yang Zhang}{
\email{zy2015@mail.ustc.edu.cn}
\affiliation{
  \institution{University of Science and Technology of China}
    \country{China}
}
}

\author{Zhengyi Yang}{
\email{yangzhy@mail.ustc.edu.cn}
\affiliation{
  \institution{University of Science and Technology of China}
    \country{China}
}
}

\author{Yancheng Luo}{
\email{luoyanchen@mail.ustc.edu.cn}
\affiliation{
  \institution{University of Science and Technology of China}
    \country{China}
}
}

\author{Chong Chen}{
\email{chenchong55@huawei.com}
\affiliation{
  \institution{Huawei Inc.}
    \country{China}
}
}

\author{Fuli Feng}{
\email{fulifeng93@gmail.com}
\affiliation{
  \institution{University of Science and Technology of China}
    \country{China}
}
}

\author{Qi Tian}{
\email{tian.qi1@huawei.com}
\affiliation{
  \institution{Huawei Inc.}
    \country{China}
}
}


\renewcommand{\shortauthors}{Bao and Zhang, et al.}





\begin{abstract}
As the focus on Large Language Models (LLMs) in the field of recommendation intensifies, the optimization of LLMs for recommendation purposes (referred to as LLM4Rec) assumes a crucial role in augmenting their effectiveness in providing recommendations. 
However, existing approaches for LLM4Rec often assess performance using restricted sets of candidates, which may not accurately reflect the models' overall ranking capabilities. 
In this paper, our objective is to investigate the comprehensive ranking capacity of LLMs and propose a two-step grounding framework known as BIGRec (Bi-step Grounding Paradigm for Recommendation). 
It initially grounds LLMs to the recommendation space by fine-tuning them to generate meaningful tokens for items and subsequently identifies appropriate actual items that correspond to the generated tokens. 
By conducting extensive experiments on two datasets, we substantiate the superior performance, capacity for handling few-shot scenarios, and versatility across multiple domains exhibited by BIGRec. 
Furthermore, we observe that the marginal benefits derived from increasing the quantity of training samples are modest for BIGRec, implying that LLMs possess the limited capability to assimilate statistical information, such as popularity and collaborative filtering, due to their robust semantic priors. 
These findings also underline the efficacy of integrating diverse statistical information into the LLM4Rec framework, thereby pointing towards a potential avenue for future research.
Our code and data are available at \url{https://github.com/SAI990323/Grounding4Rec}.
\end{abstract}

\maketitle
\section{Introduction}
Large Language Models (LLMs) have garnered significant success across various domains (such as Computer Vision~\cite{xie2023towards} and Robotics~\cite{driess2023palm}) due to their enormous context comprehension and generation capabilities~\cite{chowdhery2022palm, llm_survey}. 
Surpassing traditional language models such as BERT~\cite{bert} and GPT2~\cite{gpt2}, LLMs encode more knowledge, possess mightier reasoning abilities, and can be smoothly adapted to a new task through in-context learning with a few examples~\cite{bao2023tallrec, alpaca, chatgpt_mol}.     
In light of this, exploring utilizing LLMs for recommendation (LLM4Rec) is evolving vigorously~\cite{wu2023survey, fan2023recommender, FaiRLLM, IR_report}. 
Due to the lack of recommendation training during the pre-training stage of LLM~\cite{bao2023tallrec}, 
tuning LLMs plays a crucial role in helping the LLM achieve better recommendation performance.
\begin{figure*}
    \centering
    \includegraphics[width=0.95\textwidth]{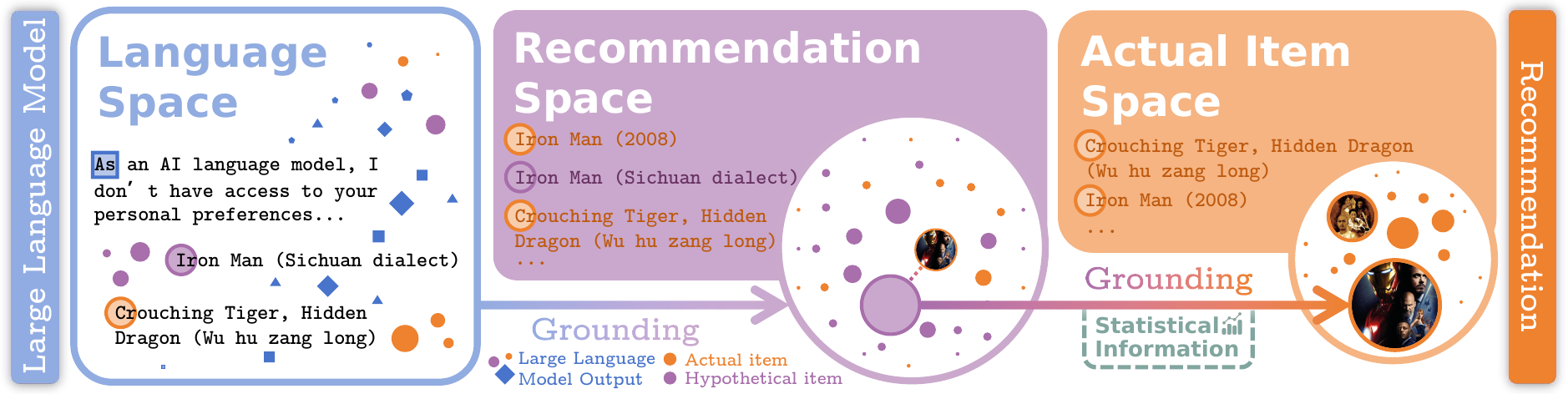}
    \caption{Illustration of the BIGRec paradigm. During the first step,  
    we ground the language space to recommendation space, which enables the model to generate token sequences of potential items including both actual and hypothetical items.
    During the second step, we ground the recommendation space to actual item space to provide users with suggestions for real-world items.
    In the second step, we can easily incorporate statistical information (\eg popularity and collaborative information) to obtain better recommendations.
    } 
    
    \label{fig:paradigm}
\end{figure*}


Numerous studies currently enhance the recommendation performance of LLMs through instruction-based fine-tuning techniques and utilizing recommendation data, achieving promising results~\cite{bao2023tallrec, zhang2023recommendation}.
When evaluating the effectiveness of these fine-tuning methods, 
they often only conduct recommendation predictions on a limited candidate set (\eg click-through rate prediction~\cite{bao2023tallrec} or negative sampling setting~\cite{zhang2023recommendation}), which do not take into account the model's overall ability to rank globally.
Interestingly, there is a study that proposes the avoidance of evaluating recommendation models through sampling, which may lead to a poor indicator of the true performance of recommender systems~\cite{krichene2020sampled}.
Therefore, we aspire to investigate and explore the capabilities of LLMs in the realm of all-rank recommendation sorting.


To fully leverage the recommendation capabilities of LLMs in an all-rank scenario, the LLM-based recommendation model must satisfy three requirements. 
1)  It must be efficient to rank a considerable number of items for a user. 
2) It should provide enough flexibility to generate a meaningful item to utilize the generation and comprehension abilities of LLMs.
%
3) It is important to consider recommending an actual item that exists in the real world while also incorporating relevant statistical information, such as popularity and collaborative information.

In light of this, we should first make the LLM consistent between the instruction tuning phase and the pre-training phase of LLMs in a generative manner.
Simultaneously, we should make the recommendation compatible with the creation of meaningful items that do not actually exist in the real world (\eg Iron Man (Sichuan dialect)\footnote{The Sichuan dialect is the prevalent vernacular in a region of China. Although the movie may not be included in the recommended list or might not even exist in reality, it could still be considered appropriate for individuals hailing from Sichuan.}), as LLMs are capable of generating creative content. 
Lastly, it is imperative to consider statistical information from past user behaviors when making recommendations to enhance their usefulness.



%
%

To achieve this goal, we formulate a \underline{bi}-step \underline{g}rounding paradigm for \underline{rec}ommendation (BIGRec) with ``language space'' $\rightarrow$ ``recommendation space'' $\rightarrow$ ``actual item space'' as shown in Figure~\ref{fig:paradigm}. 
The language space refers to the set of all possible sequences that can be generated by the LLM; the recommendation space is a subset of the language space that includes descriptions of various items satisfying user preferences, which encompasses both actual and hypothetical items.
In the first step, we ground LLMs to the recommendation space by fine-tuning them to generate meaningful tokens for item descriptions (see Table~\ref{tab:example}). 
The second step involves identifying the most suitable actual items that match the generated tokens, utilizing their latent representations obtained from the LLM. 
This step also offers flexibility for incorporating diverse statistical information required for recommendations, such as weighting the distance of representations according to item popularity.

We conduct extensive experiments on two real-world datasets to investigate the extraordinary ability of BIGRec and the influence of popularity and collaborative information.
The results first demonstrate the strong few-shot and multi-domain abilities of BIGRec, which largely outperforms traditional recommendation models and existing LLM4Rec methods. 
Remarkably, BIGRec can outperform most of the traditional models trained with 100 or even 1,000 times more samples.
Furthermore, we observe that the benefits obtained from scaling up training samples are relatively modest for BIGRec when compared to traditional models.
Considering that more training samples benefit traditional models with richer statistical information, we postulate that LLMs may take limited statistical information (\eg popularity and collaborative information) from training samples due to their powerful semantic priors. 
We further validate the effectiveness of incorporating popularity and collaborative information, showing the potential generalization of the paradigm for future works.

In conclusion, our contributions are as follows:
\begin{itemize}[leftmargin=*]
    \item We study LLM4Rec in an all-rank setting and introduce a bi-step grounding paradigm, which leverages the comprehension and generation ability of LLMs in an efficient and effective manner with the support of seamlessly integrating statistical information.

    \item We validate the effectiveness of BIGRec, which shows extraordinary ability for few-shot and cross-domain recommendation; and reveal the influence of scaling up the training data.

    \item We integrate popularity and collaborative information into BIGRec and reveal the benefit of that statistical information in LLM4Rec, showing potential future directions.
    %
    
\end{itemize}

\section{Related Work}
\paragraph{\textbf{LLM-based Recommendation}}
Researchers in the field primarily investigate LLM-based recommendations from two perspectives. Firstly, they aim to leverage the well-established in-context learning (ICL) approach within the natural language processing (NLP) community~\cite{liu2023chatgpt, friedman2023leveraging, chen2023palr}.
LLMs like ChatGPT can learn the recommendation task through autoregression by providing appropriate instructions and a few examples. 
However, previous studies have explored the effectiveness of ICL-based recommendations and have observed that the performance of these methods is often limited, which is attributed to the constraints imposed on the size of input for LLMs and the lack of recommendation knowledge~\cite{liu2023chatgpt,bao2023tallrec}.
To overcome this problem, the other perspective believes that we should apply the LLM to the recommendation task by fine-tuning the LLM due to the lack of relevant recommendation data in the pre-training phase of the LLM~\cite{bao2023tallrec, li2023exploring}.
Among them, \citep{li2023exploring} uses the LLM to get the item embedding which severs as an item representation and is fed into the traditional recommendation models (e.g. SASRec~\cite{SASRec}).
Nevertheless, this method still relies on traditional models making it difficult to take advantage of the generative ability of LLMs.
In addition, two existing studies, TALLRec~\cite{bao2023tallrec} and InstructRec~\cite{zhang2023recommendation}, employ recommendation data in an instruction-tuning phase to improve the recommendation capability of the LLM.
Yet, albeit they do consider the use of LLM from a generative standpoint, their evaluation is limited to the CTR scenario or conducted through negative sampling and does not explore the LLM's ability to rank all items.

\paragraph{\textbf{Sequential Recommendation}}
The recommendation paradigm that is closest to our experimental setup is the sequential recommendation, which requires the recommender system to predict the next item a user is likely to like based on the user's historical sequence of interactions~\cite{seq_survey_1, seq_survey_2}.
Previous work has separately endeavored RNN-based models~\cite{GRU4Rec, seq_rec_1, seq_rec_2}, CNN-based models~\cite{Caser, CNN_seq_1, CNN_seq_2}, and attention-based models~\cite{SASRec, attention_seq_1, attention_seq_2} to model user behavior, garnering respectable outcomes.
Based on these methods, an increasingly large number of works are focusing on enhancing model performance through pretraining~\cite{pretrain_1, pretrain_2}, data augmentation~\cite{contrast_1, contrast_2}, debiasing techniques~\cite{debias_1, debias_2}, distributionally robust optimization~\cite{DRO1, DRO2} and other analogous methodologies.
However, constrained by conventional ID-based recommender systems, such methods lack generalization ability as well as the rapidity to adapt to novel scenarios.
Over recent years, there has been an increasing awareness of the role semantics plays in the recommendation, with efforts now being made to leverage language models to aid the recommendation process~\cite{GPT4Rec, unirec}.
However, these methods, either persist with based language models employing encoders to extract features, which requires enormous data for pre-training, or utilize word overlapping to retrieve items, which lacks utilizing semantic information and is easy to be influenced by noise.

\paragraph{\textbf{Grounding in LLM}}
Currently, in the research of LLMs, we can mainly consider the concept of grounding from two perspectives: modality grounding and affordance grouding~\cite{wang2023interactive}.
The former is in order to ground linguistic modalities to the knowledge of other modalities, such as images, audio, or other modalities, through this approach, one can enable LLMs to capture and process richer information from the real world~\cite{multimodelgrounding}.
In the line of the modality grounding~\cite{zeng, li2020oscar, ju2022prompting}, individuals endeavor at introducing information of other modalities during both the training and inference phases of LLMs, leading to the capability of processing multimodal inputs, as exemplified by Vicuna~\cite{vicuna, vicuna2}, MiniGPT4~\cite{zhu2023minigpt}.
However due to the requirements of a large amount of data and resources~\cite{wang2023interactive}, within this paper, we have not accorded precedence to this grounding strategy but leave it as a future work.
The latter aims at grounding the LLM to a specific contextual scenario, ensuring that the results generated by the model are associated with the task rather than detached from the scenario~\cite{affordancegrouding}.
In practice, people achieve this goal via instruction-tuning on domain-specific data or utilizing customized prompts to steer the LLM~\cite{wang2023interactive}.
Where in the recommender domain, owing to its gap with the generative task itself -- we necessitate recommending an item that truly exists, hence we apply a step grounded in the generation output to the real world to fulfill the recommendation.

\section{BIGRec}
In this section, we introduce an elementary implementation of BIGRec with two grounding steps.

\subsection{Preliminary}

\paragraph{\textbf{Defination}}
To better understand the grounding paradigm, we first give definitions of the following key phrases:
\begin{itemize}[leftmargin=*]
    \item \textbf{Language Space.} This space prior to the grounding paradigm encompasses all conceivable language sequences that an LLM could generate, such as the statement, ``As an AI language model, I don't have access to your personal preferences ...''. It is not feasible to utilize this space directly for generating recommendations due to its vast and varied nature.
    \item \textbf{Recommendation Space.} 
    This is a sub-space within the language space that includes a wide range of entities that fulfill the user's preferences. These entities can represent both actual and imaginary items in a particular domain. 
    However, it is important to note that recommending purely imaginary entities may not be appropriate. 
    For instance, suggesting ``Iron Man (Sichuan dialect)'' as a recommendation would be infeasible.
    
    \item \textbf{Actual Item Space.} 
    The actual item space contains only the actual items in the recommendation space. 
    Recommending items from this actual item space is necessary. 
    For example, in the context of movie recommendations, the recommended items must be selected from the available movies on the platform.
    
\end{itemize}


To fine-tune LLMs for recommendation, we propose the BIGRec paradigm with two grounding steps. 
Firstly, we ground the output of LLMs from the language space to the recommendation space for a specific recommendation task. 
Secondly, we ground it from the recommendation space to the actual item space, enabling the recommendations of actual items to users. 
To demonstrate the capabilities of our paradigm, we present a simple implementation, which illustrates the potential and possibilities of this BIGRec paradigm.

\subsection{Implementation}
\begin{table}
\centering
  \caption{
        Example of the instruction-tuning data for the step of grounding to the space.
    }
 \label{tab:example}
    \begin{tabularx}{0.45\textwidth}{lX}    
    \toprule
        \multicolumn{2}{c}{\textbf{Instruction Input}} \\
        \cdashline{1-2}[1pt/2.5pt]\noalign{\vskip 0.5ex} 
        \textbf{Instruction:}  & Given ten movies that the user watched recently, please recommend a new movie that the user likes to the user.\\ 
        \cdashline{1-2}[1pt/2.5pt]\noalign{\vskip 0.5ex} 
        
        \textbf{Input:} & The user has watched the following movies before: ``Traffic (2000)'', ``Ocean's Eleven (2001)'', ... ``Fargo (1996)'' \\  
        \midrule
        \multicolumn{2}{c}{\textbf{Instruction Output}}  \\
        \cdashline{1-2}[1pt/2.5pt]\noalign{\vskip 0.5ex}
        \textbf{Output:}  & ``Crouching Tiger, Hidden Dragon (Wu hu zang long) (2000)''  \\ 
    \bottomrule
    \end{tabularx}
\vspace{-0.1cm}
\end{table}

In this subsection, we describe how we implement the BIGRec paradigm for recommendations. 
\subsubsection{Step 1: Grounding Language Space to Recommendation Space}
In accordance with the methodology proposed in \citep{bao2023tallrec}, we perform an instruction-tuning phase on the alpaca self-instruct data~\cite{alpaca} using LLaMA~\cite{LLaMA}.
Afterward, we conduct a recommendation-specific instruction-tuning to restrict the output of LLMs from the language space to the recommendation space. 
As demonstrated in Table \ref{tab:example}, we fine-tune LLMs in a generative manner: given a user's past interactions with items, we ask LLMs to generate a new item as the recommendation to the user. 
By fine-tuning with such data, we limit the LLMs' output to the designated recommendation space as instructed.
However, due to the creativity of the LLM, it is hard to ensure that the output of the LLMs will correspond to an actual item that exists in the real world. 
Therefore, it is essential to ground the output of LLMs to the actual item space.

\subsubsection{Step 2: Grounding Recommendation Space to Actual ItemsSpace}
In this subsection, we elaborate on how to anchor the recommender space to the actual item space.
Firstly, we align the output of LLMs with real-world items based on the representations of LLMs to implement a vanilla version of BIGRec. 
Then, we introduce statistical information (\eg popularity and collaborative information) to accurately locate the actual items for recommendations.
Specifically, we 
extract the latent representation of the generated tokens and the embedding of the actual items. Thereafter, we rank these actual items by calculating the L2 distance between their embeddings. 
The L2 distance is obtained as follows:
\begin{small}
\begin{equation}
\begin{split}
    D_i &= ||\textbf{emb}_i - \textbf{oracle}||_2,
    \label{eq:eq0}
\end{split}
\end{equation}
\end{small}
where $\textbf{emb}_i$ denotes the embedding of the $i$-th item and $\textbf{oracle}$ denotes the embedding of the outputs generated by the LLM.

\paragraph{\textbf{Injection of Statistical Information}}
We then introduce how we incorporate popularity information and collaborative information into the grounding step.
To inject popularity, we follow the idea in PDA~\cite{PDA} and 
reweight the L2 distance in Eq. (\ref{eq:eq0}) by popularity. 
In detail, we first calculate the popularity factor of each item from the following equation:
\begin{small}
\begin{equation}
\left \{
\begin{aligned}
    C_i &= \frac{\mathcal{N}^i}{\sum_{j \in \mathcal{I}} \mathcal{N}^j}, \\ 
    P_i &= \frac{C_i - \min_{j \in \mathcal{I}} \{C_j\}}{\max_{j \in \mathcal{I}} \{C_j\} - \min_{j \in \mathcal{I}} \{C_j\}},
    \label{eq:eq1}
\end{aligned}
\right .
\end{equation}
\end{small}
where $\mathcal{N}$ denotes the set of user-item interactions in the training data, $\mathcal{N}^j$ denotes the number of observed interactions for item $j$ in $\mathcal{N}$ and $\mathcal{I}$ denotes all items, $C_i$ represents the popularity of the $i$-th item, and $P_i$ is the normalized value of $C_i$. 

We then adjust the L2 distance in Eq. (\ref{eq:eq0}) by popularity:
\begin{small}
\begin{equation}
\left \{
\begin{aligned}
    \hat{D}_i &= \frac{D_i - \min_{j \in \mathcal{I}} \{D_j\}}{\max_{j \in \mathcal{I}} \{D_j\} - \min_{j \in \mathcal{I}} \{D_j\}}, \\
    \widetilde{D}_i &= \frac{\hat{D}_i}{(1 + P_i) ^ \gamma},
    \label{eq:eq2}
\end{aligned}
\right .
\end{equation}
\end{small}
where $D_i$ denotes the L2 distance between the embedding of $i$-th item and the embedding of the generated outputs by LLMs, $\hat{D}_i$ is the normalized $D_i$, and $\widetilde{D}_i$ reweights $\hat{D}_i$ by using popularity. 
To reweight $\hat{D}_i$, we utilize the inverse popularity and introduce a hyperparameter $\gamma$ to regulate the influence of popularity. 
By putting popularity in the denominator, a popular item will be assigned a smaller L2 distance and a higher rank. 
This slightly differs from the original implementation in PDA, which directly reweights prediction scores instead of L2 distance\footnote{In Section~\S\ref{sec:RQ5}, we still reweight prediction scores using PDA~\cite{PDA} to incorporate popularity into two traditional models.}.

Different from popularity, it is non-trivial to use statistical methods to quantify collaborative information. 
Considering that traditional recommendation models rely on collaborative filtering for recommendations, we regard the prediction score by these collaborative filtering models as collaborative information. 
Akin to the injection of popularity, we can substitute the variable $P_i$ in Eq.~(\ref{eq:eq2}) with the prediction score to inject collaborative information into the grounding process.
For more details, please refer to Section~\S\ref{sec:RQ5}.


\section{Experiments}
\label{section:exp}
In this section, we conduct experiments to answer the following research questions: 
\begin{itemize}[leftmargin=*]
\item RQ1: 
How does the performance of BIGRec compare to existing methods, when trained on a limited sample of 1024 data points?
\item RQ2: 
How does the performance of our model trained with limited data compare to that of traditional 
 recommendation models trained with $100\times$ or even $1,000\times$ more samples?
\item RQ3:
Can BIGRec achieve significant performance gains with increasing training data like traditional models?
\item RQ4: 
To what extent can the performance of BIGRec be enhanced by integrating popularity/collaborative information during model grounding?

\end{itemize}

\subsection{Experimental Settings}
\subsubsection{Datasets.} We conduct experiments on two datasets:
\begin{itemize}[leftmargin=*]
    \item \textbf{Movie.} This refers to the well-known benchmark dataset for movie recommendation --- MovieLens10M\footnote{https://grouplens.org/datasets/movielens/10m/}. It contains 10,682 items, 10,000,054 interactions, and 9,301,274 interaction sequences.
    
    \item \textbf{Game.} This is a video-games recommendation dataset from Amazon\footnote{https://jmcauley.ucsd.edu/data/amazon/} and we use its 5-core subset. The dataset contains 17,408 items, 496,315 interactions, and 149,796 interaction sequences.
\end{itemize}

\begin{figure}
    \centering
    \tiny
    \begin{subfigure}{0.45\textwidth}
        \centering
        \includegraphics[width=7cm, height=3.5cm]{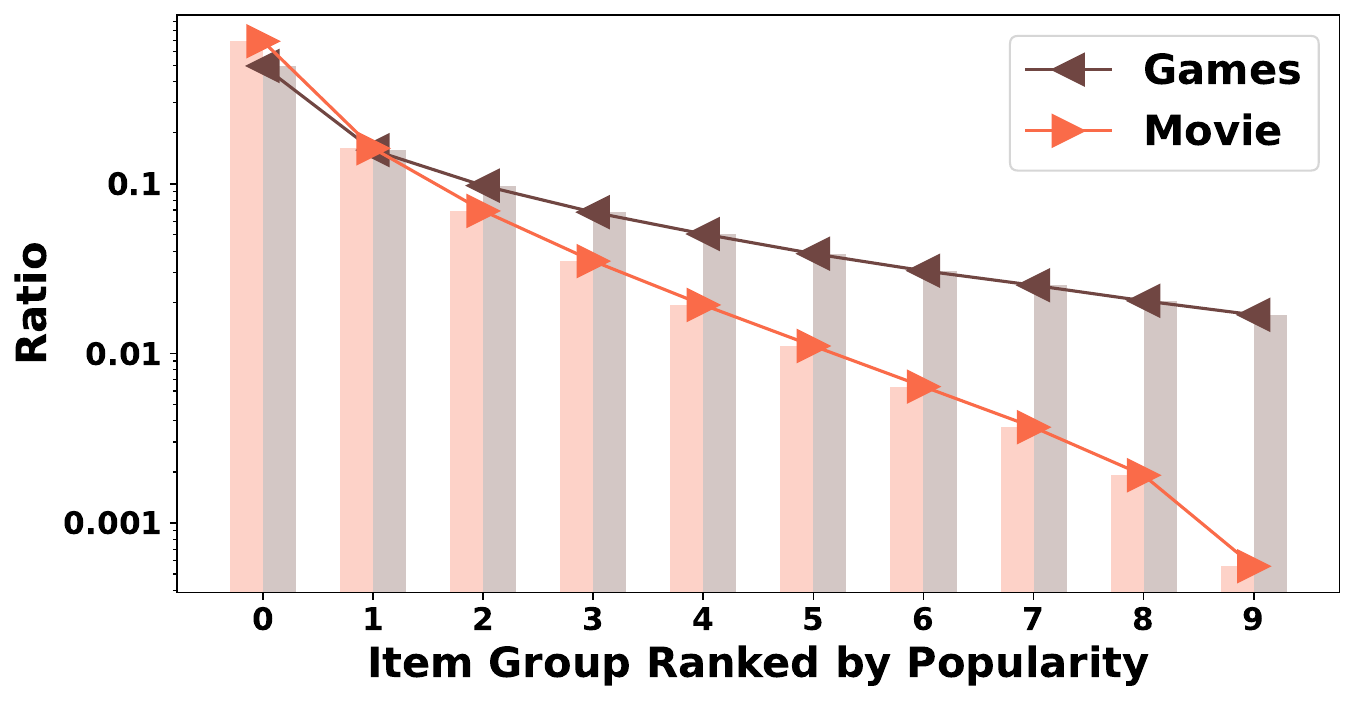}
        \caption{The distribution of items with varying popularity. It is evident that the left item groups have higher levels of popularity in this figure.}
        \label{fig:pop_count}
    \end{subfigure}
    \hfill
    \begin{subfigure}{0.45\textwidth}
        \centering
        \includegraphics[width=7cm, height=3.5cm]{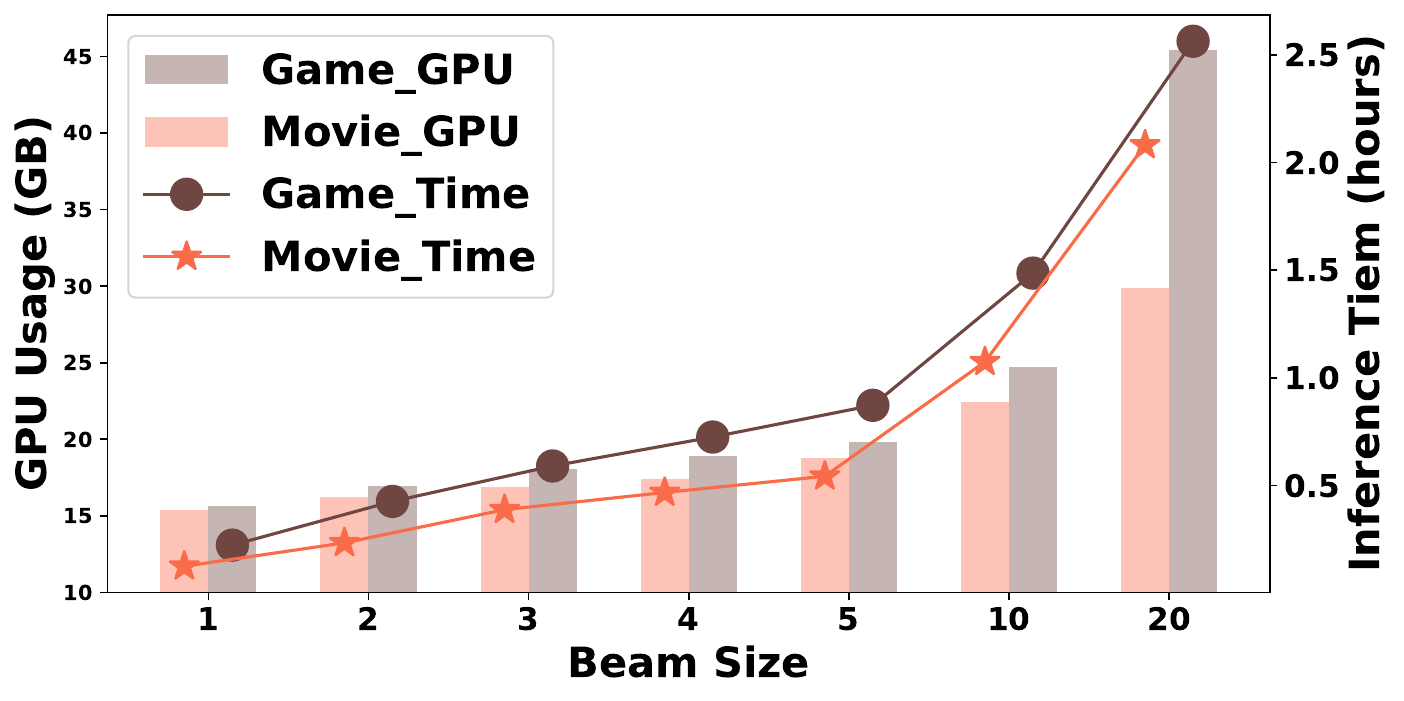}
        \caption{The GPU usage and inference time for different beam sizes with fp16 model and batch size equals to 1 for fair comparision.}
        \label{fig:beam}
    \end{subfigure}
    \caption{The two figure show the distribution of items with different popularity and the GPU usage and inference time for different beam sizes on two dataset, respectively. }
    \vspace{-0.5cm}

\end{figure}

We chose the two datasets intentionally, as they exhibit different characteristics with regard to popularity bias. 
The Movie dataset reveals a significant bias toward popular items, while the Game dataset demonstrates a lesser degree of bias. 
Figure~\ref{fig:pop_count} illustrates the frequency of interactions across items with different levels of popularity, confirming that the Game dataset presents a more equitable distribution of interactions between popular and unpopular items than the Movie dataset. 
This suggests a lower degree of popularity bias in the Game dataset. 

To simulate real-world sequential recommendation scenarios, we divided each dataset into 10 periods based on the timestamp of the interactions. Subsequently, we split the periods of each dataset into training, validation, and testing sets\footnote{Due to limitations in the inference speed of LLM models, we randomly sample 5,000 validation and testing interactions as the final validation and testing sets, respectively.} using a ratio of 8:1:1. This approach ensures that the predicted interactions during testing come after all interactions observed during training, preventing the leakage of the testing phase information~\cite{ji2023critical} during training model. This is more akin to real-world situations~\cite{refCTR,ji2023critical}.


\subsubsection{Evaluation Protocols.}

Following the previous work~\cite{DRO1, SASRec}, for each testing interaction, we set the input historical interaction sequence as the set of interactions that happened immediately prior to it. This approach allows for the possibility of including interactions that occur during the testing phase in the input sequence. 
To evaluate the model's performance, we use two commonly used evaluation metrics: Hit Ratio (HR) and Normalized Discounted Cumulative Gain (NDCG), which we compute using the all-ranking protocol~\cite{DRO1}. 
In this protocol, all items that a user has not interacted with are considered potential candidates.

\subsubsection{Compared Method}
To demonstrate the superiority of our method in the sequential recommendation, we compare it against the following conventional methods and LLM-based  methods:
\begin{itemize}[leftmargin=*]
    \item[-] \textbf{GRU4Rec~\cite{GRU4Rec}.} This is an RNN-based model that uses Gated Recurrent Units (GRU) to encode user's past interactions with items and generate recommendations based on the learned patterns.
     \item[-] \textbf{Caser~\cite{Caser}.} It is a CNN-based model that represents the sequence of recent items as an image and uses convolutional filters in both horizontal and vertical directions to capture sequential patterns.
    In our implementation, we use one vertical filter and 16 horizontal filters with searching the height in \{2, 3, 4\}.
    \item[-] \textbf{SASRec~\cite{SASRec}.} This is a self-attention-based model that uses a causal (left-to-right) attention mechanism to learn sequential patterns and predict the next item. 
    \item[-] \textbf{P5~\cite{P5}}. This approach utilizes the T5~\cite{T5} model as the backbone model and leverages a combination of item IDs and natural language templates to undergo continued pre-training for various recommendation tasks\footnote{
    For a fair comparison, we exclusively retain the task of sequence recommendation and its corresponding multiple templates during the training process.
    }.
  \item[-]\textbf{DROS~\cite{DRO1}}. This is the state-of-the-art sequential recommendation method that employs distributionally robust optimization techniques to enhance recommendation robustness against distributional changes.   
     \item[-] \textbf{ GPT4Rec-LLaMA~\cite{GPT4Rec}} is a modified version of the language model-based recommendation approach, GPT4Rec. The original method uses GPT-2 to generate hypothetical "search queries" based on a user's historical data, which are then searched using the BM25 search engine\footnote{We use the Rank-BM25 package at \url{https://github.com/dorianbrown/rank_bm25}.} to retrieve recommended items 
    .
    To ensure a fair comparison, we have replaced the GPT-2 model with the LLaMA-7B model and performed model tuning using the LoRA~\cite{LoRA} technique.   
\end{itemize}

We primarily utilize decoder-only LLMs as the backbone LLM model in our \textbf{BIGRec} method, given their prominent roles in the LLMs field. 
Specifically, we have chosen LLaMA-7B as our default choice for this study. 
In addition to GPT4Rec, there are other existing LLM-based recommendation methods, such as TALLRec~\cite{bao2023tallrec}. 
However, since these methods are not applicable to our all-ranking setting, we do not compare them in our study.

\subsubsection{Implementation Details}
We implement all of our methods using PyTorch. 
To preprocess the data, we first pad user interaction history with lengths less than 11 to a fixed length of 11. 
Then, we use a sliding window of length 11 to extract sequences, where the last item in each sequence is used as the prediction target, and the preceding items from the past interaction sequence for it. 
For all conventional models, we optimize them using the binary cross-entropy loss and uniformly sample negative samples. We use the Adam~\cite{Adam} as the optimizer with a tuned learning rate of $1e-3$, a batch size of 1024, and tune the weight decay in the range of $[1e\text{-}2, 1e\text{-}3, 1e\text{-}4, 1e\text{-}5, 1e\text{-}6, 1e\text{-}7]$. 
Regarding the hyperparameters of the conventional model architectures, we set their embedding size to 64 and the dropout ratio to 0.1. 
As suggested by the original papers, we only use one GRU layer for GRU4Rec; 
for SASRec, we set the number of self-attention heads and blocks to 1. 
We implemented DROS based on SASRec, as it performs the best in most scenarios among the three conventional sequential recommendation methods. 
For LLM-based methods, we follow the Alpaca setting~\cite{alpaca}, directly set the learning rate to $1e-4$, and use the AdamW~\cite{AdamW} optimizer.
During the generation, due to the substantial computational and inferential costs associated with the extensive use of GPU shown in Figure~\ref{fig:beam}, we follow the setting of the previous work~\cite{chowdhery2022palm, attentionallyouneed, GPT3} and employ a beam size of 4 to generate meaningful outputs. 

For the hyperparameters of BM25, we follow the tuning range provided in GPT4Rec~\cite{GPT4Rec} paper. 
For model selection, we employ an early stop strategy with a patience of 20 epochs for baseline methods and 5 epochs for LLM-based methods. Additionally, we report the average results for three random seeds.
For the hyperparameters of $\gamma$ in Eq.~(\ref{eq:eq2}), due to the disparate calculation methods employed by traditional models and LLM-based recommendation models, when incorporating information, we conducted a detailed search for a balanced integration of both approaches\footnote{Range from 0 to 1 with an increment of 0.01, and from 1 to 100 with an increment of 1.}.

\begin{table*}[]
\centering
\caption{
Comparison of model performances under the few-shot training setting (1024 samples) with NDCG@K (NG@K) and HR@K (HR@K) as the metrics. The best results are highlighted in bold, and the sub-optimal results are underlined. `Improve' refers to the relative improvement of BIGRec over the best baseline.
}
\label{tab:pop_1024}
\resizebox{0.95\textwidth}{!}{
\begin{tabular}{cccccccccccc}
\hline
Dataset                        & Model                                        & NG@1                                        & NG@3                                        & NG@5                                       & NG@10                                      & NG@20                                      & HR@1                                        & HR@3                                        & HR@5                                       & HR@10                                      & HR@20                                      \\ \cline{1-12}
                        & GRU4Rec                                      & 0.0015                                     & 0.0034                                     & 0.0047                                    & 0.0070                                    & 0.0104                                    & 0.0015                                     & 0.0047                                     & 0.0079                                    & 0.0147                                    & 0.0281                                    \\
                        & Caser                                        & 0.0020                                     & 0.0035                                     & 0.0052                                    & {0.0078}                                    & {0.0109}                                    & 0.0020                                     & 0.0046                                     & 0.0088                                    & {0.0171}                                    & {0.0293}                                    \\ 
                        & SASRec                                       & 0.0023                                     & {0.0051}                                     & {0.0062}                                    & 0.0082                                    & 0.0117                                    & 0.0023                                     & 0.0070                                     & 0.0097                                    & 0.0161                                    & 0.0301                                    \\
                        & P5 & 0.0014 &0.0026 & 0.0036 & 0.0051 &0.0069 & 0.0014 &0.0035 & 0.0059 & 0.0107 & 0.0176 \\
                        & DROS & 0.0022 &	0.0040 &	0.0052 &	0.0081 &	0.0112 &	0.0022 &	0.0051 &	0.0081 &	0.0173 &	0.0297 \\
                        & GPT4Rec-LLaMA & 0.0016 & 0.0022 & 0.0024 & 0.0028 & 0.0035 & 0.0016 & 0.0026 & 0.0030 & 0.0044 & 0.0074 \\
                        \cline{2-12} 
                        \rowcolor[HTML]{EFEFEF}
                        \cellcolor[HTML]{FFFFFF} & \textbf{BIGRec (1024)} & \textbf{0.0176}	& \textbf{0.0214} & \textbf{0.0230} &	\textbf{0.0257} & 	\textbf{0.0283}&	\textbf{0.0176}&	\textbf{0.0241}&	\textbf{0.0281}&	\textbf{0.0366}&	\textbf{0.0471} \\
\cellcolor[HTML]{FFFFFF}\multirow{-8}{*}{Movie}    & \cellcolor[HTML]{D0D0D0}\textbf{Improve}     & \cellcolor[HTML]{D0D0D0}\textbf{654.29\%}  & \cellcolor[HTML]{D0D0D0}\textbf{323.31\%}  & \cellcolor[HTML]{D0D0D0}\textbf{273.70\%} & \cellcolor[HTML]{D0D0D0}\textbf{213.71\%} & \cellcolor[HTML]{D0D0D0}\textbf{142.55\%} & \cellcolor[HTML]{D0D0D0}\textbf{654.29\%}  & \cellcolor[HTML]{D0D0D0}\textbf{244.71\%}  & \cellcolor[HTML]{D0D0D0}\textbf{188.39\%} & \cellcolor[HTML]{D0D0D0}\textbf{111.97\%} & \cellcolor[HTML]{D0D0D0}\textbf{56.55\%}  \\ 

\hline
                        & GRU4Rec                                      & {0.0013}                                     & {0.0016}                                     & 0.0018                                   & {0.0024}                                    & 0.0030                                    & {0.0013}                                     & {0.0018}                                     & 0.0024                                    & 0.0041                                    & 0.0069                                    \\
                        & Caser                                        & 0.0007                                     & 0.0012                                     & {0.0019}                                    & {0.0024}                                    & {0.0035}                                    & 0.0007                                     & 0.0016                                     & {0.0032}                                    & {0.0048}                                    & {0.0092}                                    \\
                        & SASRec& 0.0009&	0.0012&	0.0015&	0.0020&	0.0025&	0.0009&	0.0015&	0.0021&	0.0037&	0.0057 \\
                        & P5& 0.0002& 	0.0005& 	0.0007& 	0.0010& 	0.0017& 	0.0002& 	0.0007 	& 0.0012 &	0.0023& 	0.0049 \\ 
                        & DROS & 0.0006&	0.0011&	0.0013&	0.0016&	0.0022&	0.0006&	0.0015&	0.0019&	0.0027&	0.0052 \\
                        & GPT4Rec-LLaMA & 0.0000 & 0.0000 & 0.0000 & 0.0001 & 0.0001 & 0.0000 & 0.0000 & 0.0000 & 0.0002 & 0.0002 \\
                        \cline{2-12} 
                        \rowcolor[HTML]{EFEFEF}
                        \cellcolor[HTML]{FFFFFF} & \textbf{BIGRec (1024)} & \textbf{0.0133} & 	\textbf{0.0169}&	\textbf{0.0189} & 	\textbf{0.0216} &	\textbf{0.0248} &	\textbf{0.0133} &	\textbf{0.0195} &	\textbf{0.0243} &	\textbf{0.0329} & 	\textbf{0.0457} \\
 \cellcolor[HTML]{FFFFFF}\multirow{-7}{*}{Game}  & \cellcolor[HTML]{D0D0D0}\textbf{Improve}     & \cellcolor[HTML]{D0D0D0}\textbf{952.63\%} & \cellcolor[HTML]{D0D0D0}\textbf{976.26\%} & \cellcolor[HTML]{D0D0D0}\textbf{888.19\%} & \cellcolor[HTML]{D0D0D0}\textbf{799.64\%} & \cellcolor[HTML]{D0D0D0}\textbf{613.76\%} & \cellcolor[HTML]{D0D0D0}\textbf{952.63\%} & \cellcolor[HTML]{D0D0D0}\textbf{985.19\%} & \cellcolor[HTML]{D0D0D0}\textbf{660.42\%} & \cellcolor[HTML]{D0D0D0}\textbf{586.11\%} & \cellcolor[HTML]{D0D0D0}\textbf{397.10\%} \\ 
   \hline
\end{tabular}}
\end{table*}

\subsection{Performance Comparison with Limited Training Data (RQ1)}
\label{sec:RQ1}
We first follow existing work~\cite{bao2023tallrec} to study the effectiveness of our method with limited training data (1024 training points) per dataset.  
We compare it with baselines trained with equivalent training samples. 
The comparison results are summarized in Table~\ref{tab:pop_1024}, where we draw the following main observations:
\begin{itemize}[leftmargin=*]
    \item When training data is limited, the conventional sequential baselines (GRU4Rec, Caser, SASRec) exhibit significantly worse performance than BIGRec implemented with LLM. These results are not surprising as these baselines rely on ID embeddings to build a recommender model, which is difficult to learn well with limited data. In contrast, BIGRec rapidly leverages the semantic knowledge of LLM obtained during the pre-training phase to achieve effective recommendations. These findings demonstrate the superiority of BIGRec over traditional recommendation models when dealing with limited training data. Furthermore, the improvement of BIGRec over these baselines in the top-ranked positions is much higher, indicating that BIGRec may tend to rank items of interest to users in higher positions. 

    \item While GPT4Rec-LLaMA is also LLM-based, it exhibits poor performance compared to BIGRec. We attribute this to the fact that BM25, which can also
        be thought of as a method to ground the LLM outputs into actual items, is not suitable for the recommendation task. BM25 is a retrieval tool designed for document-level text. However, for the two datasets used in our study, the queries in GPT4Rec (item titles)  are very short. As a result, BM25 is susceptible to the impact of low-frequency word noise, making it difficult to accurately retrieve relevant items.

    \item  The improvement of BIGRec over conventional models is significantly higher for the Game dataset compared to the Movie dataset. This difference is possibly due to the varying properties of popularity bias between the two datasets. Conventional methods tend to capture popular bias, while BIGRec is less affected by popularity bias. Therefore, conventional methods may perform better on the Movie dataset, in which popular items play more dominant roles as evidenced by Figure~\ref{fig:pop_count}.
\end{itemize}



\begin{figure}[t]
    \centering
    \begin{subfigure}{0.45\textwidth}
        \centering
        \includegraphics[width=7cm]{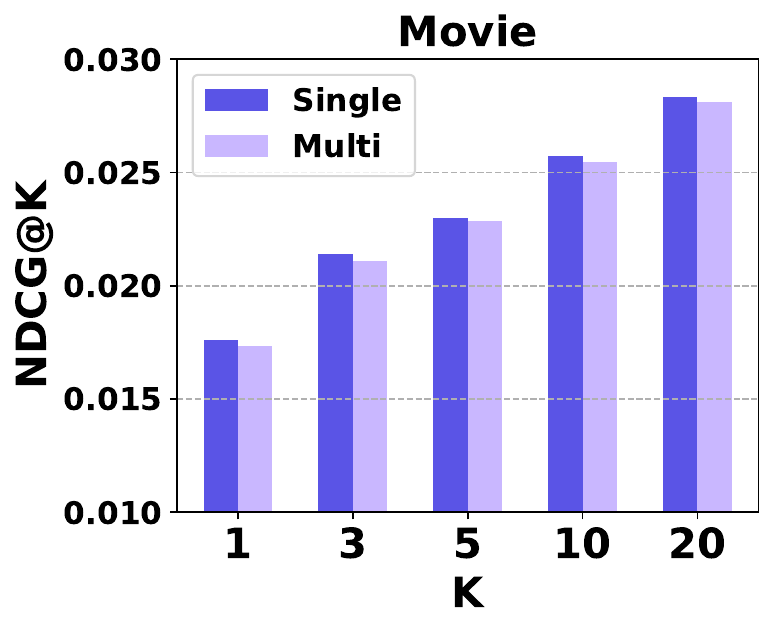}
    \end{subfigure}
    \begin{subfigure}{0.45\textwidth}
        \centering
        \includegraphics[width=7cm]{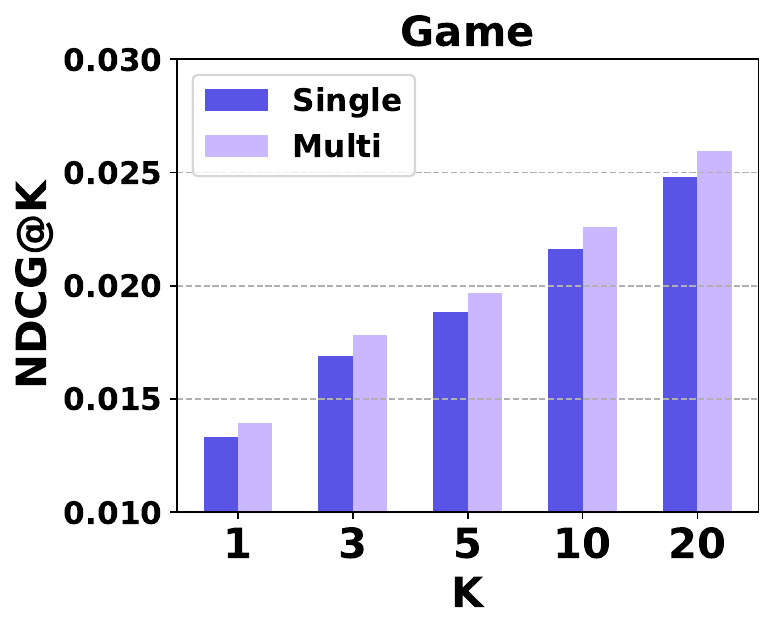}
    \end{subfigure}
    \vspace{-0.6cm}
    \caption{
    Performance comparison of BIGRec trained on multiple domain data (labeled as "Multi") and BIGRec trained on single target-domain data (labeled as "Single"), shown for NDCG@K on Movie and Game domains.
    }
    \label{fig:multi_domain_result}
\vspace{-0.4cm}
\end{figure} 

Furthermore, we also testify to the potential of BIGRec for grounding to different actual item spaces simultaneously.
To explore this, we randomly selected 1024 samples from two domains (Movie and Game) to train our model and tested it on the testing sets of two domains, respectively.
As shown in Figure~\ref{fig:multi_domain_result}, 
the results demonstrate that BIGRec trained on two-domain data performs comparably on each individual domain as compared to when it is trained on the data of the single domain. 
These findings suggest that BIGRec can simultaneously ground LLMs to different actual item spaces, at least at the level of different domains.

\subsection{Performance Comparison to Baselines Trained with More Data (RQ2)}
\label{sec:RQ3}
\begin{table*}[]
\small
\caption{
Performance comparison among baselines trained with full data, BIGRec (0) did not train the model, BIGRec (1024) trained with 1024 samples, and BIGRec (full) trained with full data.
``Most-Pop'' refers to the method recommending the most popular items. The best results are highlighted in Bold and sub-optimal results are underlined.  BIGRec (full) on Movie is omitted due to high computation cost.
}
\label{tab:full_training_result}
\begin{tabular}{cccccccccccc}
\hline

Injected                        & Model                                        & NG@1                                        & NG@3                                        & NG@5                                       & NG@10                                      & NG@20                                      & HR@1                                        & HR@3                                        & HR@5                                       & HR@10                                      & HR@20                                      \\ \hline
\multicolumn{12}{c}{\textbf{Movie}} \\
\cdashline{1-12}
                       & Most-Pop                  & 0.0032                           & 0.0076                           & 0.0088                           & 0.0121                            & 0.0170                            & 0.0032                           & 0.0108                           & 0.0138                           & 0.0244                            & 0.0438                            \\
\rowcolor[HTML]{FFFFFF} 
\cellcolor[HTML]{FFFFFF}                        & GRU4Rec              & 0.0047                 & 0.0108                           & 0.0151                           & 0.0237                            & 0.0351                            & {{0.0047}}                     & 0.0155                           & 0.0259                           & 0.0527                            & 0.0985                            \\
\rowcolor[HTML]{FFFFFF} 
\cellcolor[HTML]{FFFFFF}                        & Caser                & 0.0045                           & 0.0113                           & 0.0161                           & 0.0242                            & {{0.0354}}             & 0.0045                           & 0.0165                           & 0.0281                           & {{0.0537}}                      & {{0.0986}}             \\
\rowcolor[HTML]{FFFFFF} 
\cellcolor[HTML]{FFFFFF}                        & SASRec               & 0.0045                           & {{0.0119}}                     & {{0.0171}}                     & 0.0268                   & 0.0389                   & 0.0045                           & {{0.0175}}                     & 0.0302                  & 0.0606                   & 0.1088                   \\
\multirow{-4}{*}{\cellcolor[HTML]{FFFFFF} \shortstack{Single \\ Model \\ (+None)}}& DROS & 0.0087	& 0.0186	& 0.0245	& 0.0359	& 0.0493	& 0.0087	& 0.0261	& 0.0406	& 0.0761	& 0.1292 \\
\rowcolor[HTML]{EFEFEF}
\cellcolor[HTML]{FFFFFF} & \textbf{BIGRec (0)} & 0.0020 &	0.0034 &	0.0046&	0.0060&	0.0095&	0.0020&	0.0044&	0.0074&	0.0120&	0.0258 \\
\rowcolor[HTML]{EFEFEF}
\cellcolor[HTML]{FFFFFF}                        & \textbf{BIGRec (1024)} & \textbf{0.0176} &	\underline{0.0214} &	0.0230 &	0.0257 &	0.0283 &	\textbf{0.0176} &	0.0241& 	0.0281 &	0.0366 &	0.0471 \\ 
\cdashline{1-12}
\cdashline{1-12}
& Caser & 0.0087 & 0.0183 & 	\underline{0.0247} & 	0.0354 &	\underline{0.0494} & 	0.0087 & 	0.0258 & 	0.0404 & 	\underline{0.0756} & 	0.1296 \\
& SASRec & \underline{0.0089} & 	0.0184 & 	0.0245 & 	\underline{0.0357} & 	{0.0493} & 	\underline{0.0089} & 	\underline{0.0256} & 	\underline{0.0409} & 	0.0754 & 	\underline{0.1307} \\
\rowcolor[HTML]{EFEFEF}
\cellcolor[HTML]{FFFFFF} 
\multirow{-3}{*}{\cellcolor[HTML]{FFFFFF}+ DROS} & \textbf{BIGRec (1024)} & \textbf{0.0176}&	\textbf{0.0250}&	\textbf{0.0315}&	\textbf{0.0427}&	\textbf{0.0562}&	\textbf{0.0176}&	\textbf{0.0308}&	\textbf{0.0464}&	\textbf{0.0813}&	\textbf{0.1353} \\
\hline
\multicolumn{12}{c}{\textbf{Game}} \\
\cdashline{1-12} 
                      & Most-Pop                  & 0.0000                           & 0.0000                           & 0.0000                           & 0.0004                            & 0.0018                            & 0.0000                           & 0.0000                           & 0.0000                           & 0.0014                            & 0.0068                            \\

                        & GRU4Rec              & 0.0051                           & 0.0080                           & 0.0094                           & 0.0109                            & 0.0129                            & 0.0051                           & 0.0101                           & 0.0135                           & 0.0184                            & 0.0263                            \\
 
                       & Caser                & 0.0059                           & 0.0094                           & 0.0111                           & 0.0141                            & 0.0177                            & 0.0059                           & 0.0119                           & 0.0161                           & 0.0256                            & {{0.0401}}                      \\

                       & SASRec               & {{0.0113}}                     & {{0.0151}}                     & {{0.0164}}                     & {{0.0185}}                      & {{0.0204}}                      & {{0.0113}}                     & {{0.0179}}                     & {{0.0209}}                     & {{0.0275}}                      & 0.0353                            \\ 
\cellcolor[HTML]{FFFFFF}                        & P5              & 0.0094	& 0.0116	& 0.0131	& 0.0145 & 	0.0167	& 0.0094	& 0.0134 & 0.0172	& 0.0216	& 0.0300 \\
\cellcolor[HTML]{FFFFFF}                        & DROS &  0.0156	& 0.0194	& 0.0213	& 0.0244	& 0.0278	& 0.0156	& 0.0221	& 0.0269	& 0.0365	& 0.0500 \\
\rowcolor[HTML]{EFEFEF}
\cellcolor[HTML]{FFFFFF} & \textbf{BIGRec (0)} & 0.0020 &	0.0032&	0.0036&	0.0054&	0.0073&	0.002&	0.004&	0.005&	0.0106&	0.0184 \\
\rowcolor[HTML]{EFEFEF}
\cellcolor[HTML]{FFFFFF}                        & \textbf{BIGRec (1024)} & 0.0133 &	0.0169 &	0.0189 &	0.0216 &	0.0248 &	0.0133 &	0.0195 &	0.0243 &	0.0329 &	0.0457            \\
\rowcolor[HTML]{EFEFEF}
\cellcolor[HTML]{FFFFFF}           
\multirow{-9}{*}{\cellcolor[HTML]{FFFFFF} \shortstack{Single \\ Model \\ (+None)}} & \textbf{BIGRec (full)} & \textbf{0.0221} &	\underline{0.0250} &	\underline{0.0272} &	0.0297 &	0.0319 &	\textbf{0.0221}&	0.0270&	0.0326&	0.0401&	0.0490  \\
\cdashline{1-12} 
& Caser &\underline{0.0159}	& 0.0199 &	0.0217 &	0.0249 &	0.0288 &	\underline{0.0159} &	0.0227 &	0.0279 &	0.0375 &	0.0529\\
& SASRec & 0.0151 &	0.0197 &	0.0216 &	0.0247 &	0.0279 &	0.0151 &	0.0226 & 	0.0279 &	0.0373 &	0.0488\\
\rowcolor[HTML]{EFEFEF}
\cellcolor[HTML]{FFFFFF} 
 & \textbf{BIGRec (1024)} & 0.0133 &	0.0243 &	0.0268 &	\underline{0.0302} &	\underline{0.0338} &	0.0133 &	\underline{0.0320 }&	\underline{0.0377} &	\underline{0.0474} &	\textbf{0.0619} \\
\rowcolor[HTML]{EFEFEF}
\cellcolor[HTML]{FFFFFF} 
 \multirow{-4}{*}{\cellcolor[HTML]{FFFFFF}+ DROS} & \textbf{BIGRec (full)} &\textbf{0.0221}&	\textbf{0.0292}&	\textbf{0.0312}&	\textbf{0.0338}&	\textbf{0.0375}&	\textbf{0.0221}&	\textbf{0.0340}&	\textbf{0.0387}&	\textbf{0.0478}&	\underline{0.0611}
\\ \hline
\end{tabular}
\end{table*}
In light of the remarkable performance improvements demonstrated by BIGRec over baselines trained on limited data, we further comprehend the disparity between BIGRec trained on limited data and baselines trained with significantly larger amounts of data (100 or even 1,000 times more) to explore the extent of this gap. 

Table~\ref{tab:full_training_result} contains the results of the baselines trained with more data, which consisted of over 7 million sequences for Movie and 120 thousand for Game, alongside the results of BIGRec trained with only 1024 samples\footnote{Due to the constraints of computational resources, we only present the results of baselines that consume significant computational resources on the Game dataset.}. 
From the table, we find that, compared to conventional baselines (excluding DROS) on the Movie data, BIGRec exhibits better performance at the top position of the recommendation list, but performs worse with regard to the relative bottom positions of the recommendation list. 
Contrary to the results obtained from the Movie dataset, our method consistently demonstrates superior performance compared to traditional models (excluding DROS) when applied to the Game data. 
Compared to the meticulously optimized and state-of-the-art traditional method DROS, BIGRec trained with only $1024$ samples could still show comparable performance in most cases.
Based on the results, we draw two conclusions: 
\begin{itemize}[leftmargin=*]
%
    \item Our method, BIGRec, which was trained on a relatively small subset of the dataset with only 1,024 samples, has shown comparable performance to conventional models that were trained with a significantly larger number of data, with 100 or even 1,000 times bigger, particularly when recommendation (exposure) resources are limited. These findings highlight the practicality of utilizing LLMs for recommendation systems. Meanwhile, the results also suggest that there is potential to further enhance BIGRec's performance by increasing the size of the training set and expanding the range of items it has encountered during grounding (see BIGRec (full) in Tabel~\ref{tab:full_training_result}).
    
    \item When compared to the Game dataset, the Movie dataset exhibits higher density and is more susceptible to popularity bias (as evidenced by the performance of the Most-Pop methods on the two datasets and Figure~\ref{fig:pop_count}). The difference in the performance gain of BIGRec between the two datasets suggests that BIGRec performs exceptionally well in scenarios where data is scarce and works effectively with less reliance on popularity bias. 


\end{itemize}

\begin{figure*}[htbp] 
\tiny
\begin{minipage}[t]{\linewidth}
\centering
\includegraphics[width=\textwidth, height=2.4in]{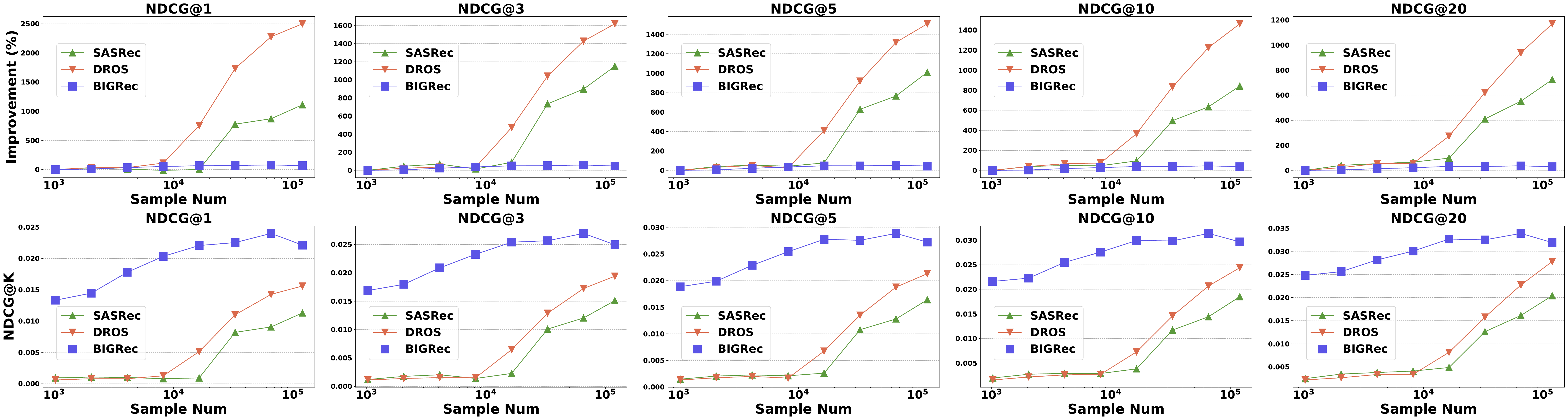}
\end{minipage}%
\caption{
Performance of SASRec, DROS, and BIGRec as training data size increases (denoted by Sample Num), along with their respective performance improvement curves relative to their initial states (1024 training samples). The bottom sub-figures showcase the recommendation performance (measured by NDCG@K), while the upper sub-figures illustrate the improvement.
}
\label{fig:scaling_law}
\end{figure*}

\subsection{Performance of BIGRec with Increasing Training Samples (RQ3)}
\label{sec:RQ5}

When comparing the results of traditional ID-based methods (such as SASRec and DROS) in Table~\ref{tab:pop_1024} and Table~\ref{tab:full_training_result}, it becomes evident that increasing the training data can enhance the model performance. This improvement can be attributed to the fact that increasing the training data aids in the acquisition of statistical information (\eg popularity and collaborative information) contained in the dataset, which is shown beneficial for recommendation~\cite{PDA, su2009survey}. 

To investigate whether LLMs can also capture dataset local statistical information in the dataset by increasing the amount of training data to enhance performance, we analyze the performance of BIGRec when increasing training data on Game. Figure~\ref{fig:scaling_law} illustrates the performance curve when increasing the training data, as well as the curve depicting the performance improvement compared to the initial state. As shown in the figure, compared to traditional recommendation models, the increase in data volume has a limited impact on the performance improvements of BIGRec. 
We guess that BIGRec is less aggressive in capturing the statistics present in the dataset, and instead prefers to utilize the semantic information of LLMs to accomplish tasks while disregarding valuable statistical information. Therefore, increasing training data could not significantly enhance BIGRec performance compared to traditional ID-based methods.


\subsection{Performance of Introducing Valuable Statistical Information (RQ4)
}

The aforementioned speculations suggest that BIGRec heavily relies on the semantic information stored in LLM while disregarding the dataset's valuable statistical information. In light of this, we believe that appropriately incorporating valuable statistical information could enhance BIGRec's recommendation capabilities. 
The consideration is that the information may be ignored by BIGRec during training but indeed describes some useful characteristics of the actual items, and thus incorporating it may enhance the grounding of BIGRec to the actual item space.
To verify it, we conduct experiments by introducing two types of statistical information: popularity and collaborative information, respectively. Note that the two types of information are recognized as valuable for recommendation systems~\cite{PDA,su2009survey}.

\begin{figure}[ht]
    \centering
    \begin{subfigure}{0.45\textwidth}
        \centering
        \includegraphics[width=7cm]{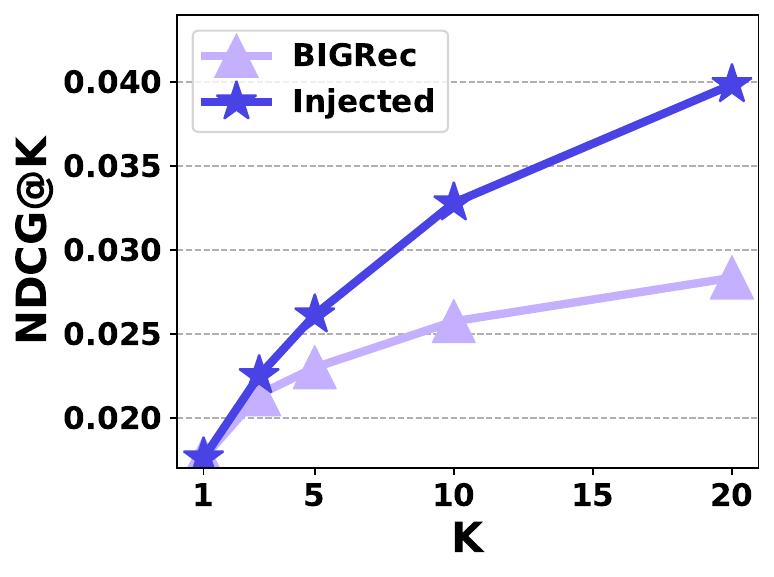}
    \end{subfigure}
    \begin{subfigure}{0.45\textwidth}
        \centering
        \includegraphics[width=7cm]{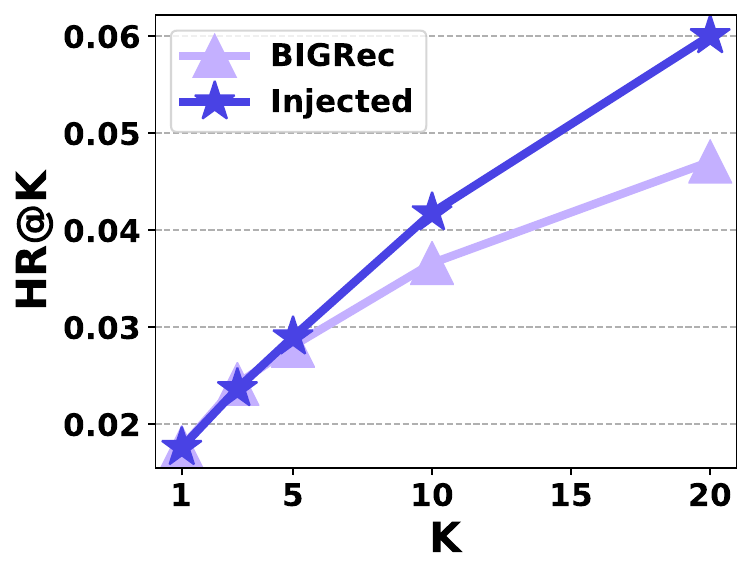}
    \end{subfigure}
    \vspace{-0.6cm}
    \caption{
    Performance comparison between BIGRec with popularity injection during grounding (labeled as ``Injected'') and the original BIGRec. NDCG@K and HR@K metrics are displayed for different values of K.
    }
    \label{fig:pop_enhanced}
\vspace{-15pt}
\end{figure}
\begin{figure}
    \centering
    \includegraphics[width=0.96\textwidth]{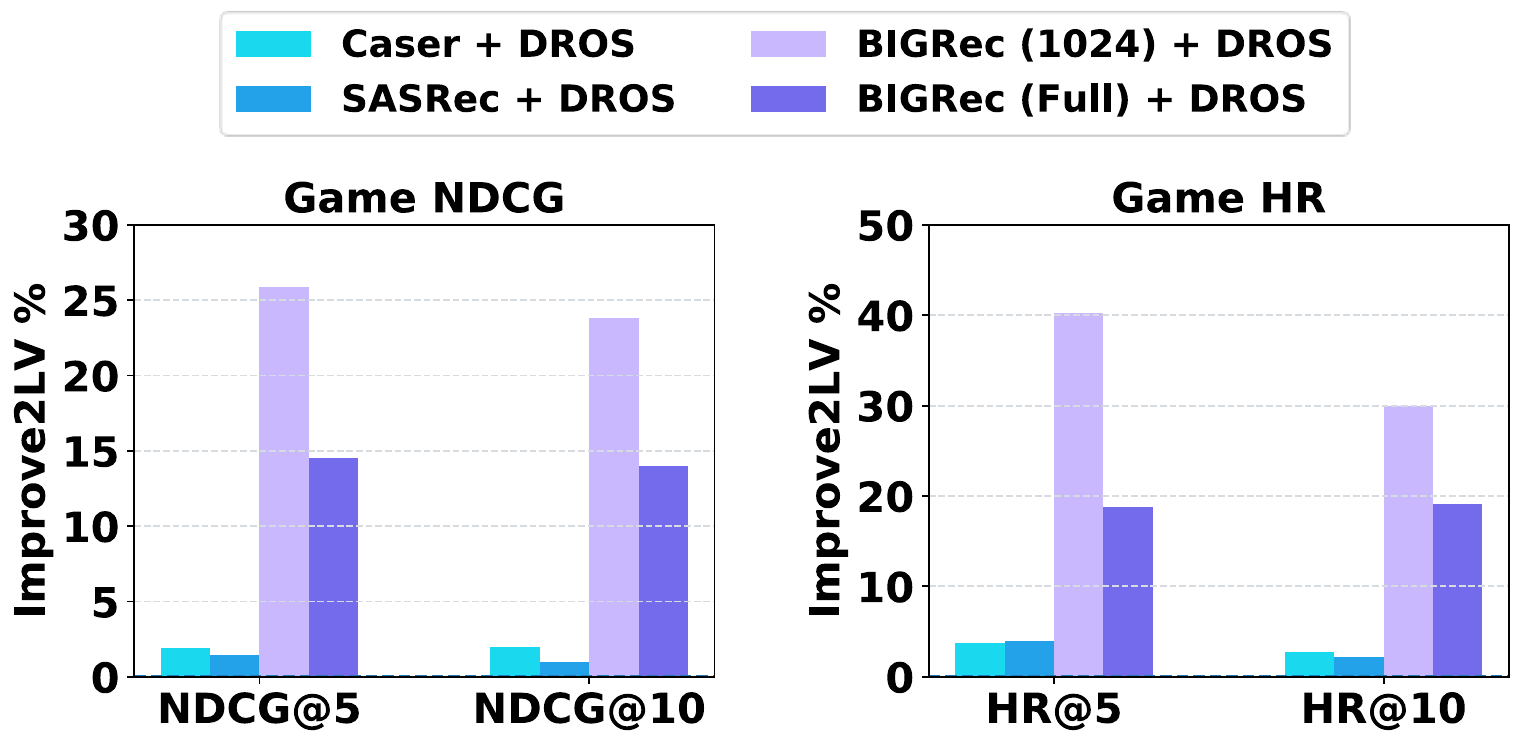}
    \vspace{-0.2cm}
    \caption{
    Performance improvement of SASRec, Caser, BIGRec (1024), and BIGRec (full) incorporated with the DROS on the Game dataset. The \underline{improve}ments are related \underline{to} the \underline{l}arger \underline{v}alue of the two combined models' performances, denoted as `Improve2LV'. 
    Notably, all the baselines here (SASRec, Caser, and DROS) are trained on the full dataset.
    } 
    
    \label{fig:improve}
    \vspace{-10pt}
\end{figure}

\noindent$\bullet$ \textbf{Introducing popularity information.} We first study the performance of BIGRec when introducing the popularity information of the training dataset using the method in Eq.~\eqref{eq:eq2}. We conduct experiments on the Movie dataset, considering that the popularity information is more important on it. We compare the original version of BIGRec with the version that incorporates popularity information in Figure~\ref{fig:pop_enhanced}. The figure demonstrates that after incorporating popularity information, BIGRec achieves performance improvements for the NDCG@K and Recall@K metrics, particularly for larger values of $K$. 
Additionally, the performance is substantially better than when using only popularity information for recommendation (as shown in Table~\ref{tab:full_training_result}). 
These results indicate that injecting popularity information can be beneficial for grounding the output of LLMs to actual items in BIGRec, further enhancing its performance.

\noindent$\bullet$ \textbf{Introducing collaborative information.} We next study the performance of BIGRec when incorporating the collaborative information encoded by conventional models (\eg DROS) into it using the method in Equation~
\eqref{eq:eq2}. For comparison, we also consider incorporating collaborative information into conventional models. The performances are shown in Tabel~\ref{tab:full_training_result}, and performance improvements are shown in Figure~\ref{fig:improve}. From the table and figure, we draw on two observations: 1) incorporating the collaborative information into both BIGRec and traditional models could bring model performance improvements; 2) incorporating collaborative information into BIGRec yields a more significant enhancement compared to incorporating information into a different conventional model. 
These results indicate that BIGRec relies more on more information (semantic information) that is different from collaborative information to generate recommendations, thus incorporating collaborative information could bring more gains.

The two experiments indicate that there is significant potential for advancing recommendation accuracy even further by investigating more effective and efficient techniques for enhancing the grounding to the actual item space in BIGRec.

        
\section{Conclusion and Future Work}
In this paper, we aim to thoroughly discuss the performance across all ranks for LLM4Rec.
In order to fully exploit the potential of LLMs, we formulate a \underline{bi}-step \underline{g}rounding paradigm for recommendation (BIGRec), which follows a generative methodology that can effectively incorporate statistical information.
Our empirical findings indicate that BIGRec trained on 1024 samples considerably outstrips that of traditional recommender models trained on an identical number of samples, attaining performance on par even with conventional models trained on the full dataset in most cases.
Furthermore, we have scaled up the training data for LLM-based recommendations. 
The results indicate that LLMs rely heavily on semantics for recommendations while ignoring some other useful information. 
Based on this insight, we have infused popularity and collaborative information into the LLM, further enhancing its capability and validating our assumption.

In view of the aforementioned discussions, it is inferred that employing a bi-step grounding approach in LLM-based recommendations shows promise. However, several unresolved issues persist.
Firstly, in terms of the initial stage of grounding in the recommendation space, our experiments reveal that satisfactory recommendation performance can be achieved with a relatively small number of samples. Besides, further increasing the sample size can potentially enhance the performance.
However, there are two opposing considerations. On one hand, training an LLM incurs considerable costs. Is it feasible to select appropriate samples for constructing the recommendation space, thereby reducing the associated expenses? On the other hand, the question of whether all items should belong to the same recommendation space remains a topic of debate.

Furthermore, concerning the second grounding step, the methodology we currently employ is rather crude, extracting embeddings using the decoder model and computing similarity is an elementary operation.
Despite our ability to regulate the production of samples with a large beam size, this approach is highly time-consuming. 
Therefore, it is imperative to find a more efficient method for connecting with actual items. 
Moreover, we have confirmed the viability of integrating item popularity and collaborative information during the grounding process. 
We hope to incorporate more conventional recommendation features more effectively to improve model performance further.


\bibliographystyle{ACM-Reference-Format}
\bibliography{AAref}


\begin{thebibliography}{61}


\ifx \showCODEN    \undefined \def \showCODEN     #1{\unskip}     \fi
\ifx \showDOI      \undefined \def \showDOI       #1{#1}\fi
\ifx \showISBNx    \undefined \def \showISBNx     #1{\unskip}     \fi
\ifx \showISBNxiii \undefined \def \showISBNxiii  #1{\unskip}     \fi
\ifx \showISSN     \undefined \def \showISSN      #1{\unskip}     \fi
\ifx \showLCCN     \undefined \def \showLCCN      #1{\unskip}     \fi
\ifx \shownote     \undefined \def \shownote      #1{#1}          \fi
\ifx \showarticletitle \undefined \def \showarticletitle #1{#1}   \fi
\ifx \showURL      \undefined \def \showURL       {\relax}        \fi
\providecommand\bibfield[2]{#2}
\providecommand\bibinfo[2]{#2}
\providecommand\natexlab[1]{#1}
\providecommand\showeprint[2][]{arXiv:#2}

\bibitem[Ahn et~al\mbox{.}(2022)]%
        {affordancegrouding}
\bibfield{author}{\bibinfo{person}{Michael Ahn}, \bibinfo{person}{Anthony
  Brohan}, \bibinfo{person}{Noah Brown}, \bibinfo{person}{Yevgen Chebotar},
  \bibinfo{person}{Omar Cortes}, \bibinfo{person}{Byron David},
  \bibinfo{person}{Chelsea Finn}, \bibinfo{person}{Chuyuan Fu},
  \bibinfo{person}{Keerthana Gopalakrishnan}, \bibinfo{person}{Karol Hausman},
  {et~al\mbox{.}}} \bibinfo{year}{2022}\natexlab{}.
\newblock \showarticletitle{Do as i can, not as i say: Grounding language in
  robotic affordances}.
\newblock \bibinfo{journal}{\emph{arXiv preprint arXiv:2204.01691}}
  (\bibinfo{year}{2022}).
\newblock


\bibitem[Ai et~al\mbox{.}(2023)]%
        {IR_report}
\bibfield{author}{\bibinfo{person}{Qingyao Ai}, \bibinfo{person}{Ting Bai},
  {and} \bibinfo{person}{et al.}} \bibinfo{year}{2023}\natexlab{}.
\newblock \showarticletitle{Information Retrieval Meets Large Language Models:
  A Strategic Report from Chinese IR Community}.
\newblock \bibinfo{journal}{\emph{arXiv preprint arXiv:2307.09751}}
  (\bibinfo{year}{2023}).
\newblock


\bibitem[Bao et~al\mbox{.}(2023)]%
        {bao2023tallrec}
\bibfield{author}{\bibinfo{person}{Keqin Bao}, \bibinfo{person}{Jizhi Zhang},
  \bibinfo{person}{Yang Zhang}, \bibinfo{person}{Wenjie Wang},
  \bibinfo{person}{Fuli Feng}, {and} \bibinfo{person}{Xiangnan He}.}
  \bibinfo{year}{2023}\natexlab{}.
\newblock \showarticletitle{Tallrec: An effective and efficient tuning
  framework to align large language model with recommendation}.
\newblock \bibinfo{journal}{\emph{Recsys short}} (\bibinfo{year}{2023}).
\newblock


\bibitem[Beinborn et~al\mbox{.}(2018)]%
        {multimodelgrounding}
\bibfield{author}{\bibinfo{person}{Lisa Beinborn}, \bibinfo{person}{Teresa
  Botschen}, {and} \bibinfo{person}{Iryna Gurevych}.}
  \bibinfo{year}{2018}\natexlab{}.
\newblock \showarticletitle{Multimodal grounding for language processing}.
\newblock \bibinfo{journal}{\emph{arXiv preprint arXiv:1806.06371}}
  (\bibinfo{year}{2018}).
\newblock


\bibitem[Brown et~al\mbox{.}(2020)]%
        {GPT3}
\bibfield{author}{\bibinfo{person}{Tom~B. Brown}, \bibinfo{person}{Benjamin
  Mann}, \bibinfo{person}{Nick Ryder}, \bibinfo{person}{Melanie Subbiah},
  \bibinfo{person}{Jared Kaplan}, \bibinfo{person}{Prafulla Dhariwal},
  \bibinfo{person}{Arvind Neelakantan}, \bibinfo{person}{Pranav Shyam},
  \bibinfo{person}{Girish Sastry}, \bibinfo{person}{Amanda Askell},
  \bibinfo{person}{Sandhini Agarwal}, \bibinfo{person}{Ariel Herbert{-}Voss},
  \bibinfo{person}{Gretchen Krueger}, \bibinfo{person}{Tom Henighan},
  \bibinfo{person}{Rewon Child}, \bibinfo{person}{Aditya Ramesh},
  \bibinfo{person}{Daniel~M. Ziegler}, \bibinfo{person}{Jeffrey Wu},
  \bibinfo{person}{Clemens Winter}, \bibinfo{person}{Christopher Hesse},
  \bibinfo{person}{Mark Chen}, \bibinfo{person}{Eric Sigler},
  \bibinfo{person}{Mateusz Litwin}, \bibinfo{person}{Scott Gray},
  \bibinfo{person}{Benjamin Chess}, \bibinfo{person}{Jack Clark},
  \bibinfo{person}{Christopher Berner}, \bibinfo{person}{Sam McCandlish},
  \bibinfo{person}{Alec Radford}, \bibinfo{person}{Ilya Sutskever}, {and}
  \bibinfo{person}{Dario Amodei}.} \bibinfo{year}{2020}\natexlab{}.
\newblock \showarticletitle{Language Models are Few-Shot Learners}. In
  \bibinfo{booktitle}{\emph{Advances in Neural Information Processing Systems
  33: Annual Conference on Neural Information Processing Systems 2020, NeurIPS
  2020, December 6-12, 2020, virtual}}, \bibfield{editor}{\bibinfo{person}{Hugo
  Larochelle}, \bibinfo{person}{Marc'Aurelio Ranzato}, \bibinfo{person}{Raia
  Hadsell}, \bibinfo{person}{Maria{-}Florina Balcan}, {and}
  \bibinfo{person}{Hsuan{-}Tien Lin}} (Eds.).
\newblock
\urldef\tempurl%
\url{https://proceedings.neurips.cc/paper/2020/hash/1457c0d6bfcb4967418bfb8ac142f64a-Abstract.html}
\showURL{%
\tempurl}


\bibitem[Chen(2023)]%
        {chen2023palr}
\bibfield{author}{\bibinfo{person}{Zheng Chen}.}
  \bibinfo{year}{2023}\natexlab{}.
\newblock \showarticletitle{PALR: Personalization Aware LLMs for
  Recommendation}.
\newblock \bibinfo{journal}{\emph{arXiv preprint arXiv:2305.07622}}
  (\bibinfo{year}{2023}).
\newblock


\bibitem[Chiang et~al\mbox{.}(2023)]%
        {vicuna}
\bibfield{author}{\bibinfo{person}{Wei-Lin Chiang}, \bibinfo{person}{Zhuohan
  Li}, \bibinfo{person}{Zi Lin}, \bibinfo{person}{Ying Sheng},
  \bibinfo{person}{Zhanghao Wu}, \bibinfo{person}{Hao Zhang},
  \bibinfo{person}{Lianmin Zheng}, \bibinfo{person}{Siyuan Zhuang},
  \bibinfo{person}{Yonghao Zhuang}, \bibinfo{person}{Joseph~E. Gonzalez},
  \bibinfo{person}{Ion Stoica}, {and} \bibinfo{person}{Eric~P. Xing}.}
  \bibinfo{year}{2023}\natexlab{}.
\newblock \bibinfo{title}{Vicuna: An Open-Source Chatbot Impressing GPT-4 with
  90\%* ChatGPT Quality}.
\newblock
\newblock
\urldef\tempurl%
\url{https://lmsys.org/blog/2023-03-30-vicuna/}
\showURL{%
\tempurl}


\bibitem[Chowdhery et~al\mbox{.}(2022)]%
        {chowdhery2022palm}
\bibfield{author}{\bibinfo{person}{Aakanksha Chowdhery},
  \bibinfo{person}{Sharan Narang}, \bibinfo{person}{Jacob Devlin},
  \bibinfo{person}{Maarten Bosma}, \bibinfo{person}{Gaurav Mishra},
  \bibinfo{person}{Adam Roberts}, \bibinfo{person}{Paul Barham},
  \bibinfo{person}{Hyung~Won Chung}, \bibinfo{person}{Charles Sutton},
  \bibinfo{person}{Sebastian Gehrmann}, {et~al\mbox{.}}}
  \bibinfo{year}{2022}\natexlab{}.
\newblock \showarticletitle{Palm: Scaling language modeling with pathways}.
\newblock \bibinfo{journal}{\emph{arXiv preprint arXiv:2204.02311}}
  (\bibinfo{year}{2022}).
\newblock


\bibitem[Cui et~al\mbox{.}(2020)]%
        {seq_rec_1}
\bibfield{author}{\bibinfo{person}{Qiang Cui}, \bibinfo{person}{Shu Wu},
  \bibinfo{person}{Qiang Liu}, \bibinfo{person}{Wen Zhong}, {and}
  \bibinfo{person}{Liang Wang}.} \bibinfo{year}{2020}\natexlab{}.
\newblock \showarticletitle{{MV-RNN:} {A} Multi-View Recurrent Neural Network
  for Sequential Recommendation}.
\newblock \bibinfo{journal}{\emph{{IEEE} Trans. Knowl. Data Eng.}}
  \bibinfo{volume}{32}, \bibinfo{number}{2} (\bibinfo{year}{2020}),
  \bibinfo{pages}{317--331}.
\newblock
\urldef\tempurl%
\url{https://doi.org/10.1109/TKDE.2018.2881260}
\showDOI{\tempurl}


\bibitem[Devlin et~al\mbox{.}(2019)]%
        {bert}
\bibfield{author}{\bibinfo{person}{Jacob Devlin}, \bibinfo{person}{Ming-Wei
  Chang}, \bibinfo{person}{Kenton Lee}, {and} \bibinfo{person}{Kristina
  Toutanova}.} \bibinfo{year}{2019}\natexlab{}.
\newblock \showarticletitle{{BERT}: Pre-training of Deep Bidirectional
  Transformers for Language Understanding}. In
  \bibinfo{booktitle}{\emph{Proceedings of the 2019 Conference of the North
  {A}merican Chapter of the Association for Computational Linguistics: Human
  Language Technologies, Volume 1 (Long and Short Papers)}}.
  \bibinfo{publisher}{Association for Computational Linguistics},
  \bibinfo{address}{Minneapolis, Minnesota}, \bibinfo{pages}{4171--4186}.
\newblock
\urldef\tempurl%
\url{https://doi.org/10.18653/v1/N19-1423}
\showDOI{\tempurl}


\bibitem[Ding et~al\mbox{.}(2022)]%
        {debias_1}
\bibfield{author}{\bibinfo{person}{Sihao Ding}, \bibinfo{person}{Fuli Feng},
  \bibinfo{person}{Xiangnan He}, \bibinfo{person}{Jinqiu Jin},
  \bibinfo{person}{Wenjie Wang}, \bibinfo{person}{Yong Liao}, {and}
  \bibinfo{person}{Yongdong Zhang}.} \bibinfo{year}{2022}\natexlab{}.
\newblock \showarticletitle{Interpolative Distillation for Unifying Biased and
  Debiased Recommendation}. In \bibinfo{booktitle}{\emph{{SIGIR} '22: The 45th
  International {ACM} {SIGIR} Conference on Research and Development in
  Information Retrieval, Madrid, Spain, July 11 - 15, 2022}},
  \bibfield{editor}{\bibinfo{person}{Enrique Amig{\'{o}}},
  \bibinfo{person}{Pablo Castells}, \bibinfo{person}{Julio Gonzalo},
  \bibinfo{person}{Ben Carterette}, \bibinfo{person}{J.~Shane Culpepper}, {and}
  \bibinfo{person}{Gabriella Kazai}} (Eds.). \bibinfo{publisher}{{ACM}},
  \bibinfo{pages}{40--49}.
\newblock
\urldef\tempurl%
\url{https://doi.org/10.1145/3477495.3532002}
\showDOI{\tempurl}


\bibitem[Donkers et~al\mbox{.}(2017)]%
        {seq_rec_2}
\bibfield{author}{\bibinfo{person}{Tim Donkers}, \bibinfo{person}{Benedikt
  Loepp}, {and} \bibinfo{person}{J{\"u}rgen Ziegler}.}
  \bibinfo{year}{2017}\natexlab{}.
\newblock \showarticletitle{Sequential user-based recurrent neural network
  recommendations}. In \bibinfo{booktitle}{\emph{Proceedings of the eleventh
  ACM conference on recommender systems}}. \bibinfo{pages}{152--160}.
\newblock


\bibitem[Driess et~al\mbox{.}(2023)]%
        {driess2023palm}
\bibfield{author}{\bibinfo{person}{Danny Driess}, \bibinfo{person}{Fei Xia},
  \bibinfo{person}{Mehdi~SM Sajjadi}, \bibinfo{person}{Corey Lynch},
  \bibinfo{person}{Aakanksha Chowdhery}, \bibinfo{person}{Brian Ichter},
  \bibinfo{person}{Ayzaan Wahid}, \bibinfo{person}{Jonathan Tompson},
  \bibinfo{person}{Quan Vuong}, \bibinfo{person}{Tianhe Yu}, {et~al\mbox{.}}}
  \bibinfo{year}{2023}\natexlab{}.
\newblock \showarticletitle{Palm-e: An embodied multimodal language model}.
\newblock \bibinfo{journal}{\emph{arXiv preprint arXiv:2303.03378}}
  (\bibinfo{year}{2023}).
\newblock


\bibitem[Fan et~al\mbox{.}(2023)]%
        {fan2023recommender}
\bibfield{author}{\bibinfo{person}{Wenqi Fan}, \bibinfo{person}{Zihuai Zhao},
  \bibinfo{person}{Jiatong Li}, \bibinfo{person}{Yunqing Liu},
  \bibinfo{person}{Xiaowei Mei}, \bibinfo{person}{Yiqi Wang},
  \bibinfo{person}{Jiliang Tang}, {and} \bibinfo{person}{Qing Li}.}
  \bibinfo{year}{2023}\natexlab{}.
\newblock \showarticletitle{Recommender Systems in the Era of Large Language
  Models (LLMs)}.
\newblock \bibinfo{journal}{\emph{arXiv preprint arXiv:2307.02046}}
  (\bibinfo{year}{2023}).
\newblock


\bibitem[Fang et~al\mbox{.}(2020)]%
        {seq_survey_1}
\bibfield{author}{\bibinfo{person}{Hui Fang}, \bibinfo{person}{Danning Zhang},
  \bibinfo{person}{Yiheng Shu}, {and} \bibinfo{person}{Guibing Guo}.}
  \bibinfo{year}{2020}\natexlab{}.
\newblock \showarticletitle{Deep Learning for Sequential Recommendation:
  Algorithms, Influential Factors, and Evaluations}.
\newblock \bibinfo{journal}{\emph{{ACM} Trans. Inf. Syst.}}
  \bibinfo{volume}{39}, \bibinfo{number}{1} (\bibinfo{year}{2020}),
  \bibinfo{pages}{10:1--10:42}.
\newblock
\urldef\tempurl%
\url{https://doi.org/10.1145/3426723}
\showDOI{\tempurl}


\bibitem[Friedman et~al\mbox{.}(2023)]%
        {friedman2023leveraging}
\bibfield{author}{\bibinfo{person}{Luke Friedman}, \bibinfo{person}{Sameer
  Ahuja}, \bibinfo{person}{David Allen}, \bibinfo{person}{Terry Tan},
  \bibinfo{person}{Hakim Sidahmed}, \bibinfo{person}{Changbo Long},
  \bibinfo{person}{Jun Xie}, \bibinfo{person}{Gabriel Schubiner},
  \bibinfo{person}{Ajay Patel}, \bibinfo{person}{Harsh Lara}, {et~al\mbox{.}}}
  \bibinfo{year}{2023}\natexlab{}.
\newblock \showarticletitle{Leveraging Large Language Models in Conversational
  Recommender Systems}.
\newblock \bibinfo{journal}{\emph{arXiv preprint arXiv:2305.07961}}
  (\bibinfo{year}{2023}).
\newblock


\bibitem[Geng et~al\mbox{.}(2022)]%
        {P5}
\bibfield{author}{\bibinfo{person}{Shijie Geng}, \bibinfo{person}{Shuchang
  Liu}, \bibinfo{person}{Zuohui Fu}, \bibinfo{person}{Yingqiang Ge}, {and}
  \bibinfo{person}{Yongfeng Zhang}.} \bibinfo{year}{2022}\natexlab{}.
\newblock \showarticletitle{Recommendation as language processing (rlp): A
  unified pretrain, personalized prompt \& predict paradigm (p5)}. In
  \bibinfo{booktitle}{\emph{Proceedings of the 16th ACM Conference on
  Recommender Systems}}. \bibinfo{pages}{299--315}.
\newblock


\bibitem[Hidasi et~al\mbox{.}(2016)]%
        {GRU4Rec}
\bibfield{author}{\bibinfo{person}{Bal{\'{a}}zs Hidasi},
  \bibinfo{person}{Alexandros Karatzoglou}, \bibinfo{person}{Linas Baltrunas},
  {and} \bibinfo{person}{Domonkos Tikk}.} \bibinfo{year}{2016}\natexlab{}.
\newblock \showarticletitle{Session-based Recommendations with Recurrent Neural
  Networks}. In \bibinfo{booktitle}{\emph{ICLR}}.
\newblock


\bibitem[Hou et~al\mbox{.}(2022)]%
        {unirec}
\bibfield{author}{\bibinfo{person}{Yupeng Hou}, \bibinfo{person}{Shanlei Mu},
  \bibinfo{person}{Wayne~Xin Zhao}, \bibinfo{person}{Yaliang Li},
  \bibinfo{person}{Bolin Ding}, {and} \bibinfo{person}{Ji-Rong Wen}.}
  \bibinfo{year}{2022}\natexlab{}.
\newblock \showarticletitle{Towards universal sequence representation learning
  for recommender systems}. In \bibinfo{booktitle}{\emph{Proceedings of the
  28th ACM SIGKDD Conference on Knowledge Discovery and Data Mining}}.
  \bibinfo{pages}{585--593}.
\newblock


\bibitem[Hu et~al\mbox{.}(2022)]%
        {LoRA}
\bibfield{author}{\bibinfo{person}{Edward~J. Hu}, \bibinfo{person}{Yelong
  Shen}, \bibinfo{person}{Phillip Wallis}, \bibinfo{person}{Zeyuan
  Allen{-}Zhu}, \bibinfo{person}{Yuanzhi Li}, \bibinfo{person}{Shean Wang},
  \bibinfo{person}{Lu Wang}, {and} \bibinfo{person}{Weizhu Chen}.}
  \bibinfo{year}{2022}\natexlab{}.
\newblock \showarticletitle{LoRA: Low-Rank Adaptation of Large Language
  Models}. In \bibinfo{booktitle}{\emph{The Tenth International Conference on
  Learning Representations, {ICLR} 2022, Virtual Event, April 25-29, 2022}}.
  \bibinfo{publisher}{OpenReview.net}.
\newblock
\urldef\tempurl%
\url{https://openreview.net/forum?id=nZeVKeeFYf9}
\showURL{%
\tempurl}


\bibitem[Ji et~al\mbox{.}(2023)]%
        {ji2023critical}
\bibfield{author}{\bibinfo{person}{Yitong Ji}, \bibinfo{person}{Aixin Sun},
  \bibinfo{person}{Jie Zhang}, {and} \bibinfo{person}{Chenliang Li}.}
  \bibinfo{year}{2023}\natexlab{}.
\newblock \showarticletitle{A critical study on data leakage in recommender
  system offline evaluation}.
\newblock \bibinfo{journal}{\emph{ACM Transactions on Information Systems}}
  \bibinfo{volume}{41}, \bibinfo{number}{3} (\bibinfo{year}{2023}),
  \bibinfo{pages}{1--27}.
\newblock


\bibitem[Ju et~al\mbox{.}(2022)]%
        {ju2022prompting}
\bibfield{author}{\bibinfo{person}{Chen Ju}, \bibinfo{person}{Tengda Han},
  \bibinfo{person}{Kunhao Zheng}, \bibinfo{person}{Ya Zhang}, {and}
  \bibinfo{person}{Weidi Xie}.} \bibinfo{year}{2022}\natexlab{}.
\newblock \showarticletitle{Prompting visual-language models for efficient
  video understanding}. In \bibinfo{booktitle}{\emph{European Conference on
  Computer Vision}}. Springer, \bibinfo{pages}{105--124}.
\newblock


\bibitem[Kang and McAuley(2018)]%
        {SASRec}
\bibfield{author}{\bibinfo{person}{Wang{-}Cheng Kang} {and}
  \bibinfo{person}{Julian~J. McAuley}.} \bibinfo{year}{2018}\natexlab{}.
\newblock \showarticletitle{Self-Attentive Sequential Recommendation}. In
  \bibinfo{booktitle}{\emph{{IEEE} International Conference on Data Mining,
  {ICDM} 2018, Singapore, November 17-20, 2018}}. \bibinfo{publisher}{{IEEE}
  Computer Society}, \bibinfo{pages}{197--206}.
\newblock


\bibitem[Kingma and Ba(2015)]%
        {Adam}
\bibfield{author}{\bibinfo{person}{Diederik~P. Kingma} {and}
  \bibinfo{person}{Jimmy Ba}.} \bibinfo{year}{2015}\natexlab{}.
\newblock \showarticletitle{Adam: {A} Method for Stochastic Optimization}. In
  \bibinfo{booktitle}{\emph{3rd International Conference on Learning
  Representations, {ICLR} 2015, San Diego, CA, USA, May 7-9, 2015, Conference
  Track Proceedings}}, \bibfield{editor}{\bibinfo{person}{Yoshua Bengio} {and}
  \bibinfo{person}{Yann LeCun}} (Eds.).
\newblock
\urldef\tempurl%
\url{http://arxiv.org/abs/1412.6980}
\showURL{%
\tempurl}


\bibitem[Krichene and Rendle(2020)]%
        {krichene2020sampled}
\bibfield{author}{\bibinfo{person}{Walid Krichene} {and}
  \bibinfo{person}{Steffen Rendle}.} \bibinfo{year}{2020}\natexlab{}.
\newblock \showarticletitle{On sampled metrics for item recommendation}. In
  \bibinfo{booktitle}{\emph{Proceedings of the 26th ACM SIGKDD international
  conference on knowledge discovery \& data mining}}.
  \bibinfo{pages}{1748--1757}.
\newblock


\bibitem[Li et~al\mbox{.}(2023b)]%
        {chatgpt_mol}
\bibfield{author}{\bibinfo{person}{Jiatong Li}, \bibinfo{person}{Yunqing Liu},
  \bibinfo{person}{Wenqi Fan}, \bibinfo{person}{Xiao-Yong Wei},
  \bibinfo{person}{Hui Liu}, \bibinfo{person}{Jiliang Tang}, {and}
  \bibinfo{person}{Qing Li}.} \bibinfo{year}{2023}\natexlab{b}.
\newblock \showarticletitle{Empowering Molecule Discovery for Molecule-Caption
  Translation with Large Language Models: A ChatGPT Perspective}.
\newblock \bibinfo{journal}{\emph{arXiv preprint arXiv:2306.06615}}
  (\bibinfo{year}{2023}).
\newblock


\bibitem[Li et~al\mbox{.}(2023c)]%
        {GPT4Rec}
\bibfield{author}{\bibinfo{person}{Jinming Li}, \bibinfo{person}{Wentao Zhang},
  \bibinfo{person}{Tian Wang}, \bibinfo{person}{Guanglei Xiong},
  \bibinfo{person}{Alan Lu}, {and} \bibinfo{person}{Gerard Medioni}.}
  \bibinfo{year}{2023}\natexlab{c}.
\newblock \showarticletitle{GPT4Rec: A generative framework for personalized
  recommendation and user interests interpretation}.
\newblock \bibinfo{journal}{\emph{arXiv preprint arXiv:2304.03879}}
  (\bibinfo{year}{2023}).
\newblock


\bibitem[Li et~al\mbox{.}(2023a)]%
        {li2023exploring}
\bibfield{author}{\bibinfo{person}{Ruyu Li}, \bibinfo{person}{Wenhao Deng},
  \bibinfo{person}{Yu Cheng}, \bibinfo{person}{Zheng Yuan},
  \bibinfo{person}{Jiaqi Zhang}, {and} \bibinfo{person}{Fajie Yuan}.}
  \bibinfo{year}{2023}\natexlab{a}.
\newblock \showarticletitle{Exploring the Upper Limits of Text-Based
  Collaborative Filtering Using Large Language Models: Discoveries and
  Insights}.
\newblock \bibinfo{journal}{\emph{arXiv preprint arXiv:2305.11700}}
  (\bibinfo{year}{2023}).
\newblock


\bibitem[Li et~al\mbox{.}(2020)]%
        {li2020oscar}
\bibfield{author}{\bibinfo{person}{Xiujun Li}, \bibinfo{person}{Xi Yin},
  \bibinfo{person}{Chunyuan Li}, \bibinfo{person}{Pengchuan Zhang},
  \bibinfo{person}{Xiaowei Hu}, \bibinfo{person}{Lei Zhang},
  \bibinfo{person}{Lijuan Wang}, \bibinfo{person}{Houdong Hu},
  \bibinfo{person}{Li Dong}, \bibinfo{person}{Furu Wei}, {et~al\mbox{.}}}
  \bibinfo{year}{2020}\natexlab{}.
\newblock \showarticletitle{Oscar: Object-semantics aligned pre-training for
  vision-language tasks}. In \bibinfo{booktitle}{\emph{Computer Vision--ECCV
  2020: 16th European Conference, Glasgow, UK, August 23--28, 2020,
  Proceedings, Part XXX 16}}. Springer, \bibinfo{pages}{121--137}.
\newblock


\bibitem[Liu et~al\mbox{.}(2023)]%
        {liu2023chatgpt}
\bibfield{author}{\bibinfo{person}{Junling Liu}, \bibinfo{person}{Chao Liu},
  \bibinfo{person}{Renjie Lv}, \bibinfo{person}{Kang Zhou}, {and}
  \bibinfo{person}{Yan Zhang}.} \bibinfo{year}{2023}\natexlab{}.
\newblock \showarticletitle{Is chatgpt a good recommender? a preliminary
  study}.
\newblock \bibinfo{journal}{\emph{arXiv preprint arXiv:2304.10149}}
  (\bibinfo{year}{2023}).
\newblock


\bibitem[Loshchilov and Hutter(2019)]%
        {AdamW}
\bibfield{author}{\bibinfo{person}{Ilya Loshchilov} {and}
  \bibinfo{person}{Frank Hutter}.} \bibinfo{year}{2019}\natexlab{}.
\newblock \showarticletitle{Decoupled Weight Decay Regularization}. In
  \bibinfo{booktitle}{\emph{7th International Conference on Learning
  Representations, {ICLR} 2019, New Orleans, LA, USA, May 6-9, 2019}}.
  \bibinfo{publisher}{OpenReview.net}.
\newblock
\urldef\tempurl%
\url{https://openreview.net/forum?id=Bkg6RiCqY7}
\showURL{%
\tempurl}


\bibitem[Ni et~al\mbox{.}(2018)]%
        {pretrain_1}
\bibfield{author}{\bibinfo{person}{Yabo Ni}, \bibinfo{person}{Dan Ou},
  \bibinfo{person}{Shichen Liu}, \bibinfo{person}{Xiang Li},
  \bibinfo{person}{Wenwu Ou}, \bibinfo{person}{Anxiang Zeng}, {and}
  \bibinfo{person}{Luo Si}.} \bibinfo{year}{2018}\natexlab{}.
\newblock \showarticletitle{Perceive Your Users in Depth: Learning Universal
  User Representations from Multiple E-commerce Tasks}. In
  \bibinfo{booktitle}{\emph{Proceedings of the 24th {ACM} {SIGKDD}
  International Conference on Knowledge Discovery {\&} Data Mining, {KDD} 2018,
  London, UK, August 19-23, 2018}}, \bibfield{editor}{\bibinfo{person}{Yike
  Guo} {and} \bibinfo{person}{Faisal Farooq}} (Eds.).
  \bibinfo{publisher}{{ACM}}, \bibinfo{pages}{596--605}.
\newblock
\urldef\tempurl%
\url{https://doi.org/10.1145/3219819.3219828}
\showDOI{\tempurl}


\bibitem[Qiu et~al\mbox{.}(2021)]%
        {contrast_2}
\bibfield{author}{\bibinfo{person}{Ruihong Qiu}, \bibinfo{person}{Zi Huang},
  \bibinfo{person}{Hongzhi Yin}, {and} \bibinfo{person}{Zijian Wang}.}
  \bibinfo{year}{2021}\natexlab{}.
\newblock \showarticletitle{Contrastive Learning for Representation
  Degeneration Problem in Sequential Recommendation}.
\newblock \bibinfo{journal}{\emph{CoRR}}  \bibinfo{volume}{abs/2110.05730}
  (\bibinfo{year}{2021}).
\newblock
\showeprint[arXiv]{2110.05730}
\urldef\tempurl%
\url{https://arxiv.org/abs/2110.05730}
\showURL{%
\tempurl}


\bibitem[Qiu et~al\mbox{.}(2022)]%
        {contrast_1}
\bibfield{author}{\bibinfo{person}{Ruihong Qiu}, \bibinfo{person}{Zi Huang},
  \bibinfo{person}{Hongzhi Yin}, {and} \bibinfo{person}{Zijian Wang}.}
  \bibinfo{year}{2022}\natexlab{}.
\newblock \showarticletitle{Contrastive Learning for Representation
  Degeneration Problem in Sequential Recommendation}. In
  \bibinfo{booktitle}{\emph{{WSDM} '22: The Fifteenth {ACM} International
  Conference on Web Search and Data Mining, Virtual Event / Tempe, AZ, USA,
  February 21 - 25, 2022}}, \bibfield{editor}{\bibinfo{person}{K.~Selcuk
  Candan}, \bibinfo{person}{Huan Liu}, \bibinfo{person}{Leman Akoglu},
  \bibinfo{person}{Xin~Luna Dong}, {and} \bibinfo{person}{Jiliang Tang}}
  (Eds.). \bibinfo{publisher}{{ACM}}, \bibinfo{pages}{813--823}.
\newblock
\urldef\tempurl%
\url{https://doi.org/10.1145/3488560.3498433}
\showDOI{\tempurl}


\bibitem[Radford et~al\mbox{.}(2019)]%
        {gpt2}
\bibfield{author}{\bibinfo{person}{Alec Radford}, \bibinfo{person}{Jeff Wu},
  \bibinfo{person}{Rewon Child}, \bibinfo{person}{David Luan},
  \bibinfo{person}{Dario Amodei}, {and} \bibinfo{person}{Ilya Sutskever}.}
  \bibinfo{year}{2019}\natexlab{}.
\newblock \showarticletitle{Language Models are Unsupervised Multitask
  Learners}.
\newblock  (\bibinfo{year}{2019}).
\newblock


\bibitem[Raffel et~al\mbox{.}(2020)]%
        {T5}
\bibfield{author}{\bibinfo{person}{Colin Raffel}, \bibinfo{person}{Noam
  Shazeer}, \bibinfo{person}{Adam Roberts}, \bibinfo{person}{Katherine Lee},
  \bibinfo{person}{Sharan Narang}, \bibinfo{person}{Michael Matena},
  \bibinfo{person}{Yanqi Zhou}, \bibinfo{person}{Wei Li}, {and}
  \bibinfo{person}{Peter~J Liu}.} \bibinfo{year}{2020}\natexlab{}.
\newblock \showarticletitle{Exploring the limits of transfer learning with a
  unified text-to-text transformer}.
\newblock \bibinfo{journal}{\emph{The Journal of Machine Learning Research}}
  \bibinfo{volume}{21}, \bibinfo{number}{1} (\bibinfo{year}{2020}),
  \bibinfo{pages}{5485--5551}.
\newblock


\bibitem[Su and Khoshgoftaar(2009)]%
        {su2009survey}
\bibfield{author}{\bibinfo{person}{Xiaoyuan Su} {and} \bibinfo{person}{Taghi~M
  Khoshgoftaar}.} \bibinfo{year}{2009}\natexlab{}.
\newblock \showarticletitle{A survey of collaborative filtering techniques}.
\newblock \bibinfo{journal}{\emph{Advances in artificial intelligence}}
  \bibinfo{volume}{2009} (\bibinfo{year}{2009}).
\newblock


\bibitem[Tang and Wang(2018)]%
        {Caser}
\bibfield{author}{\bibinfo{person}{Jiaxi Tang} {and} \bibinfo{person}{Ke
  Wang}.} \bibinfo{year}{2018}\natexlab{}.
\newblock \showarticletitle{Personalized top-n sequential recommendation via
  convolutional sequence embedding}. In \bibinfo{booktitle}{\emph{WSDM}}.
  \bibinfo{pages}{565--573}.
\newblock


\bibitem[Taori et~al\mbox{.}(2023)]%
        {alpaca}
\bibfield{author}{\bibinfo{person}{Rohan Taori}, \bibinfo{person}{Ishaan
  Gulrajani}, \bibinfo{person}{Tianyi Zhang}, \bibinfo{person}{Yann Dubois},
  \bibinfo{person}{Xuechen Li}, \bibinfo{person}{Carlos Guestrin},
  \bibinfo{person}{Percy Liang}, {and} \bibinfo{person}{Tatsunori~B.
  Hashimoto}.} \bibinfo{year}{2023}\natexlab{}.
\newblock \bibinfo{title}{Stanford Alpaca: An Instruction-following LLaMA
  model}.
\newblock
  \bibinfo{howpublished}{\url{https://github.com/tatsu-lab/stanford_alpaca}}.
\newblock


\bibitem[Touvron et~al\mbox{.}(2023)]%
        {LLaMA}
\bibfield{author}{\bibinfo{person}{Hugo Touvron}, \bibinfo{person}{Thibaut
  Lavril}, \bibinfo{person}{Gautier Izacard}, \bibinfo{person}{Xavier
  Martinet}, \bibinfo{person}{Marie{-}Anne Lachaux},
  \bibinfo{person}{Timoth{\'{e}}e Lacroix}, \bibinfo{person}{Baptiste
  Rozi{\`{e}}re}, \bibinfo{person}{Naman Goyal}, \bibinfo{person}{Eric Hambro},
  \bibinfo{person}{Faisal Azhar}, \bibinfo{person}{Aur{\'{e}}lien Rodriguez},
  \bibinfo{person}{Armand Joulin}, \bibinfo{person}{Edouard Grave}, {and}
  \bibinfo{person}{Guillaume Lample}.} \bibinfo{year}{2023}\natexlab{}.
\newblock \showarticletitle{LLaMA: Open and Efficient Foundation Language
  Models}.
\newblock \bibinfo{journal}{\emph{CoRR}}  \bibinfo{volume}{abs/2302.13971}
  (\bibinfo{year}{2023}).
\newblock
\urldef\tempurl%
\url{https://doi.org/10.48550/arXiv.2302.13971}
\showDOI{\tempurl}
\showeprint[arXiv]{2302.13971}


\bibitem[Vaswani et~al\mbox{.}(2017)]%
        {attentionallyouneed}
\bibfield{author}{\bibinfo{person}{Ashish Vaswani}, \bibinfo{person}{Noam
  Shazeer}, \bibinfo{person}{Niki Parmar}, \bibinfo{person}{Jakob Uszkoreit},
  \bibinfo{person}{Llion Jones}, \bibinfo{person}{Aidan~N. Gomez},
  \bibinfo{person}{Lukasz Kaiser}, {and} \bibinfo{person}{Illia Polosukhin}.}
  \bibinfo{year}{2017}\natexlab{}.
\newblock \showarticletitle{Attention is All you Need}. In
  \bibinfo{booktitle}{\emph{Advances in Neural Information Processing Systems
  30: Annual Conference on Neural Information Processing Systems 2017, December
  4-9, 2017, Long Beach, CA, {USA}}},
  \bibfield{editor}{\bibinfo{person}{Isabelle Guyon}, \bibinfo{person}{Ulrike
  von Luxburg}, \bibinfo{person}{Samy Bengio}, \bibinfo{person}{Hanna~M.
  Wallach}, \bibinfo{person}{Rob Fergus}, \bibinfo{person}{S.~V.~N.
  Vishwanathan}, {and} \bibinfo{person}{Roman Garnett}} (Eds.).
  \bibinfo{pages}{5998--6008}.
\newblock
\urldef\tempurl%
\url{https://proceedings.neurips.cc/paper/2017/hash/3f5ee243547dee91fbd053c1c4a845aa-Abstract.html}
\showURL{%
\tempurl}


\bibitem[Wang et~al\mbox{.}(2019)]%
        {seq_survey_2}
\bibfield{author}{\bibinfo{person}{Shoujin Wang}, \bibinfo{person}{Liang Hu},
  \bibinfo{person}{Yan Wang}, \bibinfo{person}{Longbing Cao},
  \bibinfo{person}{Quan~Z. Sheng}, {and} \bibinfo{person}{Mehmet~A. Orgun}.}
  \bibinfo{year}{2019}\natexlab{}.
\newblock \showarticletitle{Sequential Recommender Systems: Challenges,
  Progress and Prospects}. In \bibinfo{booktitle}{\emph{Proceedings of the
  Twenty-Eighth International Joint Conference on Artificial Intelligence,
  {IJCAI} 2019, Macao, China, August 10-16, 2019}},
  \bibfield{editor}{\bibinfo{person}{Sarit Kraus}} (Ed.).
  \bibinfo{publisher}{ijcai.org}, \bibinfo{pages}{6332--6338}.
\newblock
\urldef\tempurl%
\url{https://doi.org/10.24963/ijcai.2019/883}
\showDOI{\tempurl}


\bibitem[Wang et~al\mbox{.}(2022)]%
        {debias_2}
\bibfield{author}{\bibinfo{person}{Zhenlei Wang}, \bibinfo{person}{Shiqi Shen},
  \bibinfo{person}{Zhipeng Wang}, \bibinfo{person}{Bo Chen},
  \bibinfo{person}{Xu Chen}, {and} \bibinfo{person}{Ji{-}Rong Wen}.}
  \bibinfo{year}{2022}\natexlab{}.
\newblock \showarticletitle{Unbiased Sequential Recommendation with Latent
  Confounders}. In \bibinfo{booktitle}{\emph{{WWW} '22: The {ACM} Web
  Conference 2022, Virtual Event, Lyon, France, April 25 - 29, 2022}},
  \bibfield{editor}{\bibinfo{person}{Fr{\'{e}}d{\'{e}}rique Laforest},
  \bibinfo{person}{Rapha{\"{e}}l Troncy}, \bibinfo{person}{Elena Simperl},
  \bibinfo{person}{Deepak Agarwal}, \bibinfo{person}{Aristides Gionis},
  \bibinfo{person}{Ivan Herman}, {and} \bibinfo{person}{Lionel M{\'{e}}dini}}
  (Eds.). \bibinfo{publisher}{{ACM}}, \bibinfo{pages}{2195--2204}.
\newblock
\urldef\tempurl%
\url{https://doi.org/10.1145/3485447.3512092}
\showDOI{\tempurl}


\bibitem[Wang et~al\mbox{.}(2023)]%
        {wang2023interactive}
\bibfield{author}{\bibinfo{person}{Zekun Wang}, \bibinfo{person}{Ge Zhang},
  \bibinfo{person}{Kexin Yang}, \bibinfo{person}{Ning Shi},
  \bibinfo{person}{Wangchunshu Zhou}, \bibinfo{person}{Shaochun Hao},
  \bibinfo{person}{Guangzheng Xiong}, \bibinfo{person}{Yizhi Li},
  \bibinfo{person}{Mong~Yuan Sim}, \bibinfo{person}{Xiuying Chen},
  {et~al\mbox{.}}} \bibinfo{year}{2023}\natexlab{}.
\newblock \showarticletitle{Interactive natural language processing}.
\newblock \bibinfo{journal}{\emph{arXiv preprint arXiv:2305.13246}}
  (\bibinfo{year}{2023}).
\newblock


\bibitem[Wen et~al\mbox{.}(2022)]%
        {DRO2}
\bibfield{author}{\bibinfo{person}{Hongyi Wen}, \bibinfo{person}{Xinyang Yi},
  \bibinfo{person}{Tiansheng Yao}, \bibinfo{person}{Jiaxi Tang},
  \bibinfo{person}{Lichan Hong}, {and} \bibinfo{person}{Ed~H Chi}.}
  \bibinfo{year}{2022}\natexlab{}.
\newblock \showarticletitle{Distributionally-robust Recommendations for
  Improving Worst-case User Experience}. In
  \bibinfo{booktitle}{\emph{Proceedings of the ACM Web Conference 2022}}.
  \bibinfo{pages}{3606--3610}.
\newblock


\bibitem[Wu et~al\mbox{.}(2023)]%
        {wu2023survey}
\bibfield{author}{\bibinfo{person}{Likang Wu}, \bibinfo{person}{Zhi Zheng},
  \bibinfo{person}{Zhaopeng Qiu}, \bibinfo{person}{Hao Wang},
  \bibinfo{person}{Hongchao Gu}, \bibinfo{person}{Tingjia Shen},
  \bibinfo{person}{Chuan Qin}, \bibinfo{person}{Chen Zhu},
  \bibinfo{person}{Hengshu Zhu}, \bibinfo{person}{Qi Liu}, {et~al\mbox{.}}}
  \bibinfo{year}{2023}\natexlab{}.
\newblock \showarticletitle{A Survey on Large Language Models for
  Recommendation}.
\newblock \bibinfo{journal}{\emph{arXiv preprint arXiv:2305.19860}}
  (\bibinfo{year}{2023}).
\newblock


\bibitem[Xie et~al\mbox{.}(2023)]%
        {xie2023towards}
\bibfield{author}{\bibinfo{person}{Lingxi Xie}, \bibinfo{person}{Longhui Wei},
  \bibinfo{person}{Xiaopeng Zhang}, \bibinfo{person}{Kaifeng Bi},
  \bibinfo{person}{Xiaotao Gu}, \bibinfo{person}{Jianlong Chang}, {and}
  \bibinfo{person}{Qi Tian}.} \bibinfo{year}{2023}\natexlab{}.
\newblock \showarticletitle{Towards AGI in Computer Vision: Lessons Learned
  from GPT and Large Language Models}.
\newblock \bibinfo{journal}{\emph{arXiv preprint arXiv:2306.08641}}
  (\bibinfo{year}{2023}).
\newblock


\bibitem[Xu et~al\mbox{.}(2021)]%
        {attention_seq_2}
\bibfield{author}{\bibinfo{person}{Chengfeng Xu}, \bibinfo{person}{Jian Feng},
  \bibinfo{person}{Pengpeng Zhao}, \bibinfo{person}{Fuzhen Zhuang},
  \bibinfo{person}{Deqing Wang}, \bibinfo{person}{Yanchi Liu}, {and}
  \bibinfo{person}{Victor~S. Sheng}.} \bibinfo{year}{2021}\natexlab{}.
\newblock \showarticletitle{Long- and short-term self-attention network for
  sequential recommendation}.
\newblock \bibinfo{journal}{\emph{Neurocomputing}}  \bibinfo{volume}{423}
  (\bibinfo{year}{2021}), \bibinfo{pages}{580--589}.
\newblock
\urldef\tempurl%
\url{https://doi.org/10.1016/j.neucom.2020.10.066}
\showDOI{\tempurl}


\bibitem[Yan et~al\mbox{.}(2019)]%
        {CNN_seq_2}
\bibfield{author}{\bibinfo{person}{An Yan}, \bibinfo{person}{Shuo Cheng},
  \bibinfo{person}{Wang{-}Cheng Kang}, \bibinfo{person}{Mengting Wan}, {and}
  \bibinfo{person}{Julian~J. McAuley}.} \bibinfo{year}{2019}\natexlab{}.
\newblock \showarticletitle{CosRec: 2D Convolutional Neural Networks for
  Sequential Recommendation}. In \bibinfo{booktitle}{\emph{Proceedings of the
  28th {ACM} International Conference on Information and Knowledge Management,
  {CIKM} 2019, Beijing, China, November 3-7, 2019}},
  \bibfield{editor}{\bibinfo{person}{Wenwu Zhu}, \bibinfo{person}{Dacheng Tao},
  \bibinfo{person}{Xueqi Cheng}, \bibinfo{person}{Peng Cui},
  \bibinfo{person}{Elke~A. Rundensteiner}, \bibinfo{person}{David Carmel},
  \bibinfo{person}{Qi~He}, {and} \bibinfo{person}{Jeffrey~Xu Yu}} (Eds.).
  \bibinfo{publisher}{{ACM}}, \bibinfo{pages}{2173--2176}.
\newblock
\urldef\tempurl%
\url{https://doi.org/10.1145/3357384.3358113}
\showDOI{\tempurl}


\bibitem[Yang et~al\mbox{.}(2023)]%
        {DRO1}
\bibfield{author}{\bibinfo{person}{Zhengyi Yang}, \bibinfo{person}{Xiangnan
  He}, \bibinfo{person}{Jizhi Zhang}, \bibinfo{person}{Jiancan Wu},
  \bibinfo{person}{Xin Xin}, \bibinfo{person}{Jiawei Chen}, {and}
  \bibinfo{person}{Xiang Wang}.} \bibinfo{year}{2023}\natexlab{}.
\newblock \showarticletitle{A Generic Learning Framework for Sequential
  Recommendation with Distribution Shifts}. In
  \bibinfo{booktitle}{\emph{Proceedings of the 46th International ACM SIGIR
  Conference on Research and Development in Information Retrieval}}.
\newblock


\bibitem[Yuan et~al\mbox{.}(2020)]%
        {pretrain_2}
\bibfield{author}{\bibinfo{person}{Fajie Yuan}, \bibinfo{person}{Xiangnan He},
  \bibinfo{person}{Alexandros Karatzoglou}, {and} \bibinfo{person}{Liguang
  Zhang}.} \bibinfo{year}{2020}\natexlab{}.
\newblock \showarticletitle{Parameter-Efficient Transfer from Sequential
  Behaviors for User Modeling and Recommendation}. In
  \bibinfo{booktitle}{\emph{Proceedings of the 43rd International {ACM} {SIGIR}
  conference on research and development in Information Retrieval, {SIGIR}
  2020, Virtual Event, China, July 25-30, 2020}},
  \bibfield{editor}{\bibinfo{person}{Jimmy~X. Huang},
  \bibinfo{person}{Yi~Chang}, \bibinfo{person}{Xueqi Cheng},
  \bibinfo{person}{Jaap Kamps}, \bibinfo{person}{Vanessa Murdock},
  \bibinfo{person}{Ji{-}Rong Wen}, {and} \bibinfo{person}{Yiqun Liu}} (Eds.).
  \bibinfo{publisher}{{ACM}}, \bibinfo{pages}{1469--1478}.
\newblock
\urldef\tempurl%
\url{https://doi.org/10.1145/3397271.3401156}
\showDOI{\tempurl}


\bibitem[Yuan et~al\mbox{.}(2019)]%
        {CNN_seq_1}
\bibfield{author}{\bibinfo{person}{Fajie Yuan}, \bibinfo{person}{Alexandros
  Karatzoglou}, \bibinfo{person}{Ioannis Arapakis}, \bibinfo{person}{Joemon~M.
  Jose}, {and} \bibinfo{person}{Xiangnan He}.} \bibinfo{year}{2019}\natexlab{}.
\newblock \showarticletitle{A Simple Convolutional Generative Network for Next
  Item Recommendation}. In \bibinfo{booktitle}{\emph{Proceedings of the Twelfth
  {ACM} International Conference on Web Search and Data Mining, {WSDM} 2019,
  Melbourne, VIC, Australia, February 11-15, 2019}},
  \bibfield{editor}{\bibinfo{person}{J.~Shane Culpepper},
  \bibinfo{person}{Alistair Moffat}, \bibinfo{person}{Paul~N. Bennett}, {and}
  \bibinfo{person}{Kristina Lerman}} (Eds.). \bibinfo{publisher}{{ACM}},
  \bibinfo{pages}{582--590}.
\newblock
\urldef\tempurl%
\url{https://doi.org/10.1145/3289600.3290975}
\showDOI{\tempurl}


\bibitem[Zeng et~al\mbox{.}(2022)]%
        {zeng}
\bibfield{author}{\bibinfo{person}{Yan Zeng}, \bibinfo{person}{Xinsong Zhang},
  \bibinfo{person}{Hang Li}, \bibinfo{person}{Jiawei Wang},
  \bibinfo{person}{Jipeng Zhang}, {and} \bibinfo{person}{Wangchunshu Zhou}.}
  \bibinfo{year}{2022}\natexlab{}.
\newblock \showarticletitle{X\({}^{\mbox{2}}\)-VLM: All-In-One Pre-trained
  Model For Vision-Language Tasks}.
\newblock \bibinfo{journal}{\emph{CoRR}}  \bibinfo{volume}{abs/2211.12402}
  (\bibinfo{year}{2022}).
\newblock
\urldef\tempurl%
\url{https://doi.org/10.48550/arXiv.2211.12402}
\showDOI{\tempurl}
\showeprint[arXiv]{2211.12402}


\bibitem[Zhang et~al\mbox{.}(2023a)]%
        {FaiRLLM}
\bibfield{author}{\bibinfo{person}{Jizhi Zhang}, \bibinfo{person}{Keqin Bao},
  \bibinfo{person}{Yang Zhang}, \bibinfo{person}{Wenjie Wang},
  \bibinfo{person}{Fuli Feng}, {and} \bibinfo{person}{Xiangnan He}.}
  \bibinfo{year}{2023}\natexlab{a}.
\newblock \showarticletitle{Is chatgpt fair for recommendation? evaluating
  fairness in large language model recommendation}.
\newblock \bibinfo{journal}{\emph{arXiv preprint arXiv:2305.07609}}
  (\bibinfo{year}{2023}).
\newblock


\bibitem[Zhang et~al\mbox{.}(2023c)]%
        {zhang2023recommendation}
\bibfield{author}{\bibinfo{person}{Junjie Zhang}, \bibinfo{person}{Ruobing
  Xie}, \bibinfo{person}{Yupeng Hou}, \bibinfo{person}{Wayne~Xin Zhao},
  \bibinfo{person}{Leyu Lin}, {and} \bibinfo{person}{Ji-Rong Wen}.}
  \bibinfo{year}{2023}\natexlab{c}.
\newblock \showarticletitle{Recommendation as instruction following: A large
  language model empowered recommendation approach}.
\newblock \bibinfo{journal}{\emph{arXiv preprint arXiv:2305.07001}}
  (\bibinfo{year}{2023}).
\newblock


\bibitem[Zhang et~al\mbox{.}(2019)]%
        {attention_seq_1}
\bibfield{author}{\bibinfo{person}{Tingting Zhang}, \bibinfo{person}{Pengpeng
  Zhao}, \bibinfo{person}{Yanchi Liu}, \bibinfo{person}{Victor~S. Sheng},
  \bibinfo{person}{Jiajie Xu}, \bibinfo{person}{Deqing Wang},
  \bibinfo{person}{Guanfeng Liu}, {and} \bibinfo{person}{Xiaofang Zhou}.}
  \bibinfo{year}{2019}\natexlab{}.
\newblock \showarticletitle{Feature-level Deeper Self-Attention Network for
  Sequential Recommendation}. In \bibinfo{booktitle}{\emph{Proceedings of the
  Twenty-Eighth International Joint Conference on Artificial Intelligence,
  {IJCAI} 2019, Macao, China, August 10-16, 2019}},
  \bibfield{editor}{\bibinfo{person}{Sarit Kraus}} (Ed.).
  \bibinfo{publisher}{ijcai.org}, \bibinfo{pages}{4320--4326}.
\newblock
\urldef\tempurl%
\url{https://doi.org/10.24963/ijcai.2019/600}
\showDOI{\tempurl}


\bibitem[Zhang et~al\mbox{.}(2021)]%
        {PDA}
\bibfield{author}{\bibinfo{person}{Yang Zhang}, \bibinfo{person}{Fuli Feng},
  \bibinfo{person}{Xiangnan He}, \bibinfo{person}{Tianxin Wei},
  \bibinfo{person}{Chonggang Song}, \bibinfo{person}{Guohui Ling}, {and}
  \bibinfo{person}{Yongdong Zhang}.} \bibinfo{year}{2021}\natexlab{}.
\newblock \showarticletitle{Causal Intervention for Leveraging Popularity Bias
  in Recommendation}. In \bibinfo{booktitle}{\emph{Proceedings of the 44th
  International ACM SIGIR Conference on Research and Development in Information
  Retrieval}} (Virtual Event, Canada) \emph{(\bibinfo{series}{SIGIR '21})}.
  \bibinfo{publisher}{Association for Computing Machinery},
  \bibinfo{address}{New York, NY, USA}, \bibinfo{pages}{11–20}.
\newblock
\showISBNx{9781450380379}
\urldef\tempurl%
\url{https://doi.org/10.1145/3404835.3462875}
\showDOI{\tempurl}


\bibitem[Zhang et~al\mbox{.}(2023b)]%
        {refCTR}
\bibfield{author}{\bibinfo{person}{Yang Zhang}, \bibinfo{person}{Tianhao Shi},
  \bibinfo{person}{Fuli Feng}, \bibinfo{person}{Wenjie Wang},
  \bibinfo{person}{Dingxian Wang}, \bibinfo{person}{Xiangnan He}, {and}
  \bibinfo{person}{Yongdong Zhang}.} \bibinfo{year}{2023}\natexlab{b}.
\newblock \showarticletitle{Reformulating {CTR} Prediction: Learning Invariant
  Feature Interactions}. In \bibinfo{booktitle}{\emph{Proceedings of the 46th
  International ACM SIGIR Conference on Research and Development in Information
  Retrieval}}.
\newblock


\bibitem[Zhao et~al\mbox{.}(2023)]%
        {llm_survey}
\bibfield{author}{\bibinfo{person}{Wayne~Xin Zhao}, \bibinfo{person}{Kun Zhou},
  \bibinfo{person}{Junyi Li}, \bibinfo{person}{Tianyi Tang},
  \bibinfo{person}{Xiaolei Wang}, \bibinfo{person}{Yupeng Hou},
  \bibinfo{person}{Yingqian Min}, \bibinfo{person}{Beichen Zhang},
  \bibinfo{person}{Junjie Zhang}, \bibinfo{person}{Zican Dong},
  \bibinfo{person}{Yifan Du}, \bibinfo{person}{Chen Yang},
  \bibinfo{person}{Yushuo Chen}, \bibinfo{person}{Zhipeng Chen},
  \bibinfo{person}{Jinhao Jiang}, \bibinfo{person}{Ruiyang Ren},
  \bibinfo{person}{Yifan Li}, \bibinfo{person}{Xinyu Tang},
  \bibinfo{person}{Zikang Liu}, \bibinfo{person}{Peiyu Liu},
  \bibinfo{person}{Jian{-}Yun Nie}, {and} \bibinfo{person}{Ji{-}Rong Wen}.}
  \bibinfo{year}{2023}\natexlab{}.
\newblock \showarticletitle{A Survey of Large Language Models}.
\newblock \bibinfo{journal}{\emph{CoRR}}  \bibinfo{volume}{abs/2303.18223}
  (\bibinfo{year}{2023}).
\newblock
\showeprint[arXiv]{2303.18223}


\bibitem[Zheng et~al\mbox{.}(2023)]%
        {vicuna2}
\bibfield{author}{\bibinfo{person}{Lianmin Zheng}, \bibinfo{person}{Wei-Lin
  Chiang}, \bibinfo{person}{Ying Sheng}, \bibinfo{person}{Siyuan Zhuang},
  \bibinfo{person}{Zhanghao Wu}, \bibinfo{person}{Yonghao Zhuang},
  \bibinfo{person}{Zi Lin}, \bibinfo{person}{Zhuohan Li},
  \bibinfo{person}{Dacheng Li}, \bibinfo{person}{Eric.~P Xing},
  \bibinfo{person}{Hao Zhang}, \bibinfo{person}{Joseph~E. Gonzalez}, {and}
  \bibinfo{person}{Ion Stoica}.} \bibinfo{year}{2023}\natexlab{}.
\newblock \bibinfo{title}{Judging LLM-as-a-judge with MT-Bench and Chatbot
  Arena}.
\newblock
\newblock
\showeprint[arxiv]{2306.05685}~[cs.CL]


\bibitem[Zhu et~al\mbox{.}(2023)]%
        {zhu2023minigpt}
\bibfield{author}{\bibinfo{person}{Deyao Zhu}, \bibinfo{person}{Jun Chen},
  \bibinfo{person}{Xiaoqian Shen}, \bibinfo{person}{Xiang Li}, {and}
  \bibinfo{person}{Mohamed Elhoseiny}.} \bibinfo{year}{2023}\natexlab{}.
\newblock \showarticletitle{MiniGPT-4: Enhancing Vision-Language Understanding
  with Advanced Large Language Models}.
\newblock \bibinfo{journal}{\emph{arXiv preprint arXiv:2304.10592}}
  (\bibinfo{year}{2023}).
\newblock


\end{thebibliography}

\end{document}